\documentclass[pdflatex,sn-mathphys-num]{sn-jnl}

\usepackage{graphicx}%
\usepackage{multirow}%
\usepackage{amsmath,amssymb,amsfonts}%
\usepackage{amsthm}%
\usepackage{mathrsfs}%
\usepackage[title]{appendix}%
\usepackage{xcolor}%
\usepackage{textcomp}%
\usepackage{manyfoot}%
\usepackage{booktabs}%
\usepackage{algorithm}%
\usepackage{algorithmicx}%
\usepackage{algpseudocode}%
\usepackage{listings}%
\usepackage{subfig}
\usepackage{tabularx}
\usepackage{amssymb}
\usepackage{array} 
\usepackage{threeparttable}
\usepackage{multicol}

\begin{document}

\title[Article Title]{Motion Exploration of Articulated Product Concepts in Interactive Sketching Environment}


\author[1]{\fnm{Kalyan Ramana} \sur{Gattoz}}\email{gkalyanramana@gmail.com, md18resch11002@iith.ac.in}

\author[2]{\fnm{Prasad S.} \sur{Onkar}}\email{psonkar@des.iith.ac.in}

\affil[1, 2]{\orgdiv{Department of Design}, \orgname{Indian Institute of Technology Hyderabad}, \orgaddress{\street{Sangareddy District}, \city{Hyderabad}, \postcode{502284}, \state{Telangana}, \country{India}}}


\abstract{In the early stages of engineering design, it is essential to know how a product behaves, especially how it moves. As designers must keep adjusting the motion until it meets the intended requirements, this process is often repetitive and time-consuming. Although the physics behind these motions is usually based on simple equations, manually working through them can be tedious and inefficient. To ease this burden, some tasks are now handled by computers. One common method involves converting hand-drawn sketches into models using CAD or CAE software. However, this approach can be time- and resource-intensive. Additionally, product sketches are usually best understood only by the designers who created them. Others may struggle to interpret them correctly, relying heavily on intuition and prior experience. Since sketches are static, they fail to show how a product moves, limiting their usefulness.
This paper presents a new approach that addresses these issues by digitising the natural act of sketching. It allows designers to create, simulate, and test the motion of mechanical concepts in a more interactive way. An application was developed to evaluate this method, focusing on user satisfaction and mental workload during a design task. The results showed a 77\% reduction in cognitive effort compared to traditional methods, with users reporting high satisfaction. Future work will focus on expanding this approach from 2D (planar) to full 3D (spatial) design environments, enabling more complex product concept development.}




\maketitle

\section*{Article highlights}
\begin{itemize}
\item Existing CAD/E software relies on primitive geometries for kinematic simulations, resulting in resource-intensive modelling processes.  
\item An integrated sketching environment was developed to facilitate motion exploration during concept sketching, providing designers with motion feedback without disrupting the sketching activity.  
\item Usability was evaluated in terms of subjective satisfaction and cognitive effort.  
\item Designers found the interface intuitive, engaging, and user-friendly, strongly favouring its adoption in their design process.  
\item The digital sketching interface led to a significant reduction in cognitive effort compared to traditional methods.
\end{itemize}
\clearpage
\section*{Abbreviations}
\begin{multicols}{2}
\begin{itemize}
  \item SIMBA: Sketching Interface for Mechanism Behaviour Analysis
  \item EU: Ease of use
  \item FU: Fun to use
  \item WU: Want to use
  \item INT: Intuitiveness
  \item F1: Sketching feature
  \item F2: Building feature
  \item F3: Attaching objects and images feature
  \item F4: Motion visualization feature
  \item F5: Path tracing feature
  \item F6: Joint manipulation feature
  \item CI: Confidence Interval
  \item SD: Standard deviation
  \item Bca: Bias corrected and accelerated
  \item IQR: Inter-quartile range
  \item Q1, Q2, Q3: First, second and third quartile
  \item IV: Independent variable
  \item DV: Dependent variable
\end{itemize}
\end{multicols}
\section{Introduction}\label{sec1}
In a typical engineering design process \cite{Pahl1996}, the conceptual design phase is one of the crucial phases because it significantly influences the overall product parameters \cite{Geoffrey2010}. It is important to support the designers in exploring the concepts from multiple perspectives to realise them without major revisions at later stages. Typically,  due to the inherent advantages of sketching in the early stages of conceptual design \cite{Jonathan1990, Prucell1998, Goldschmidt1991, Goel1991, Tversky2002}, designers use them to explore product concepts creatively. Sketches are a static representation of the mental images of designers, which helps them externalise, refine, and validate their ideas. However, some products,  called \textit{articulated products}, have relative motion between their components/parts. As reported in \cite{wetzel2009automated, pearson2019sketch}, such motion information is difficult to represent in a static sketch and less effective in supporting creative problem-solving. Hence, the designers face the challenge of mentally simulating to check whether the concept achieves the intended motion.

\begin{figure}
	\centering
 \subfloat[Anatomy of an office stapler (Source: Authors)]{\includegraphics[width=0.35\textwidth,keepaspectratio]{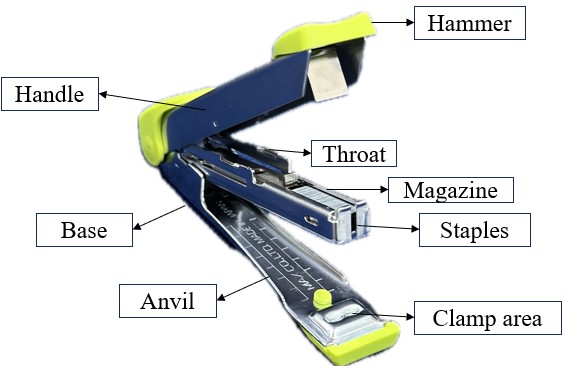}} \vspace{0.2cm}
\subfloat[Abstraction of a embodiment product to a skeleton \cite{erdam1998mechanism}]{\includegraphics[width=0.6\textwidth,keepaspectratio]{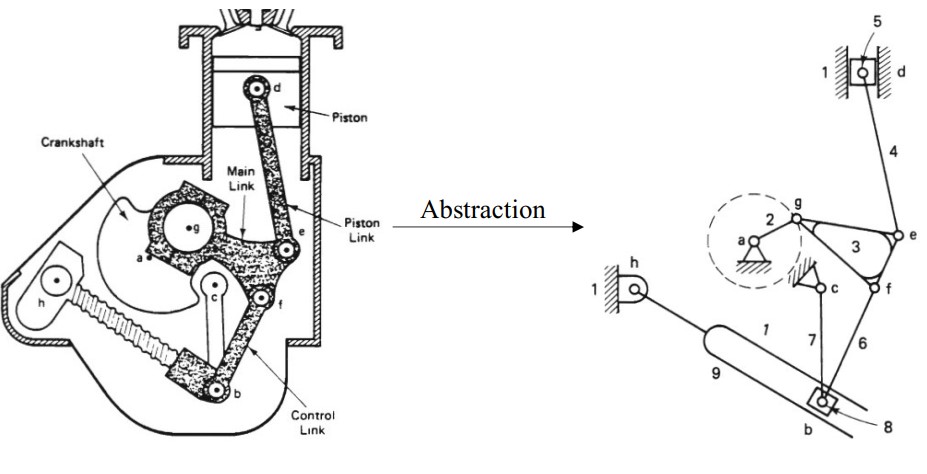}}
	\caption{Commonly observed articulated products}
	\label{fig:common_articulations}
\end{figure}

\begin{figure}[h]
\centering
     \subfloat[Subject 1]{\includegraphics[width=0.49\textwidth,keepaspectratio]{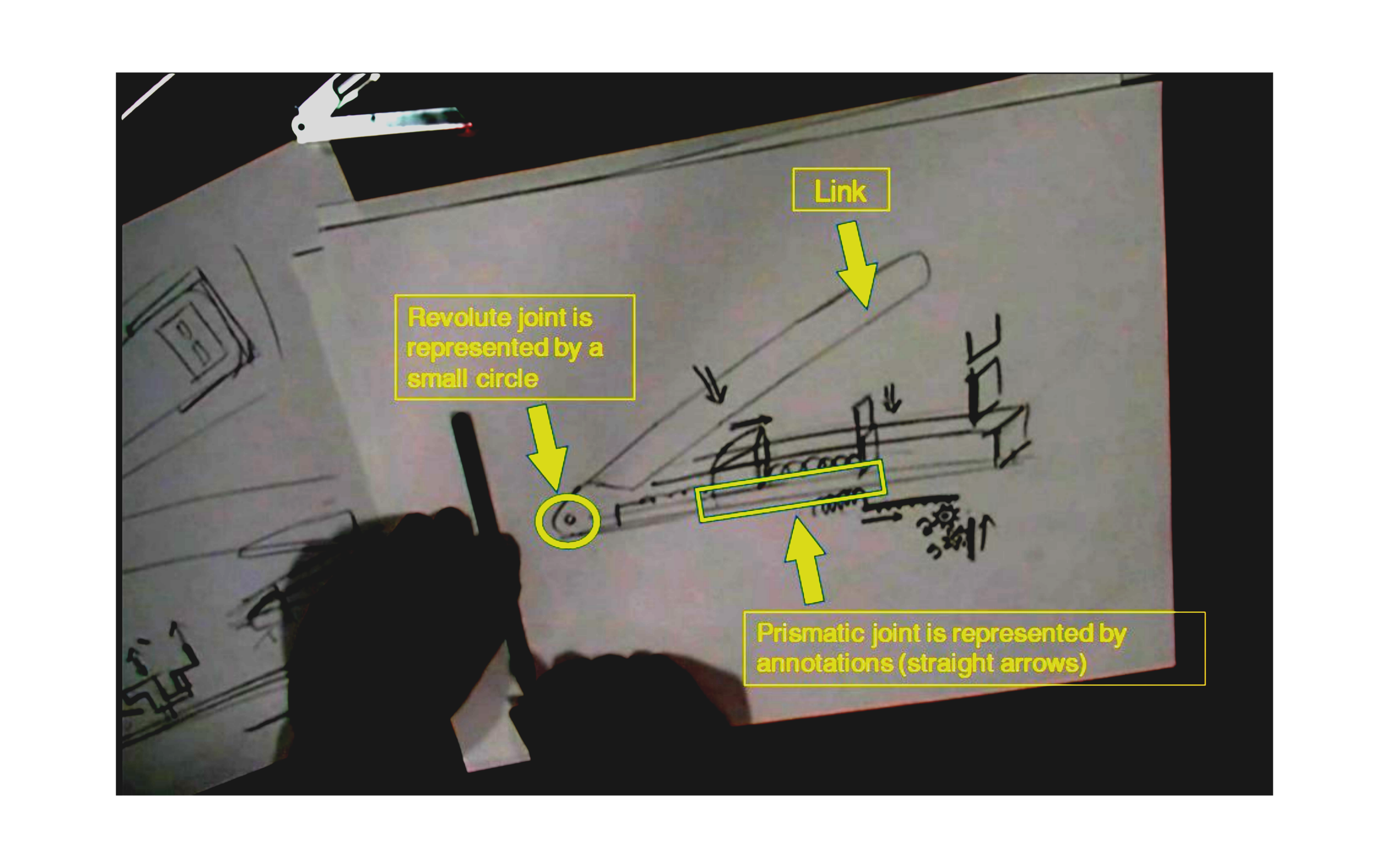}} \vspace{0.2cm}
    \subfloat[Subject 2]{\includegraphics[width=0.49\textwidth,keepaspectratio]{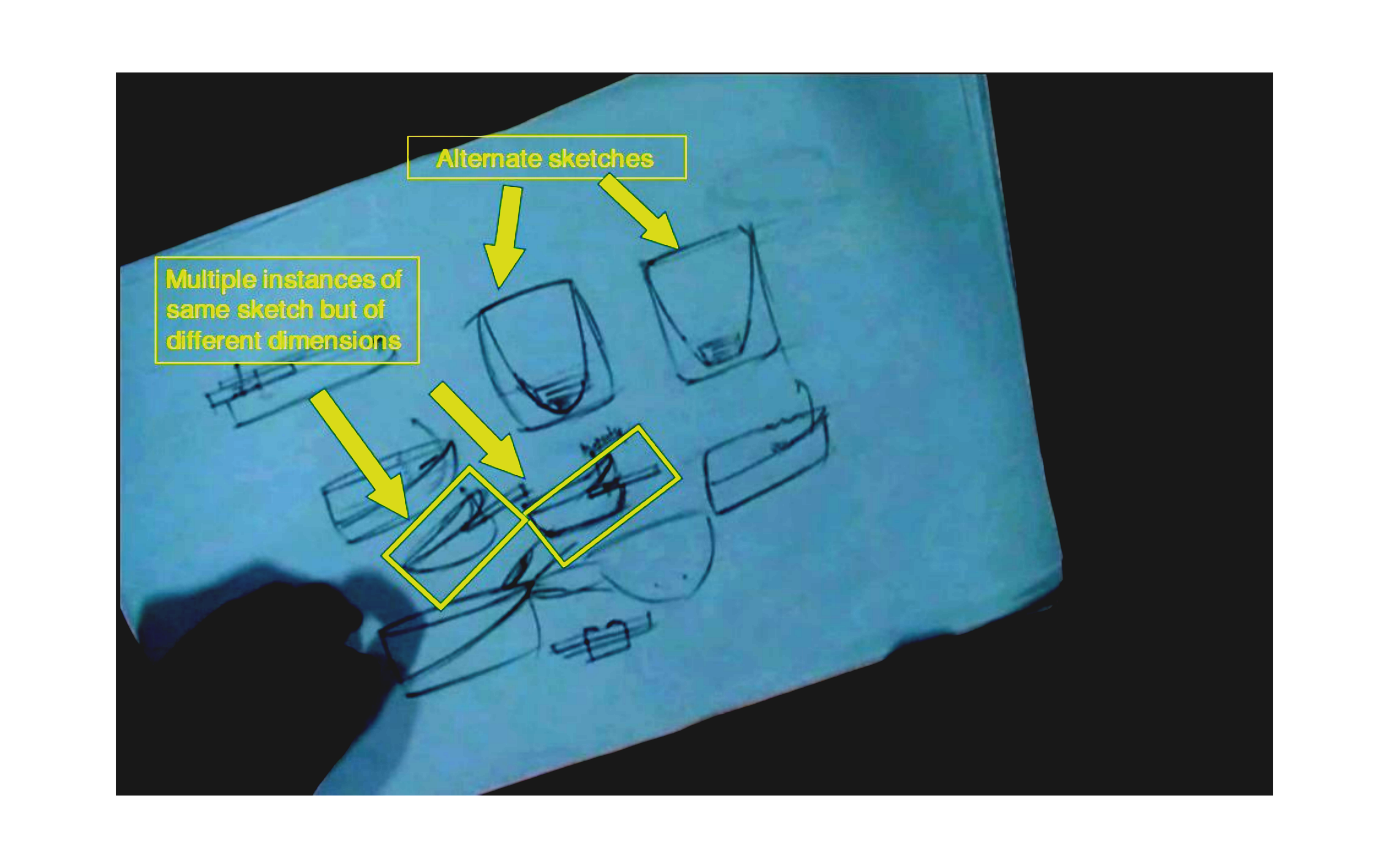}}
    \caption{Observations of traditional sketching activity during
motion exploration (Source: authors) (a) Subject 1 Sketch (b) Subject 2 Sketch.}
    \label{motion_exploration_requirements}
\end{figure}

\begin{figure}[h]
    \centering
    \includegraphics[width=0.7\linewidth]{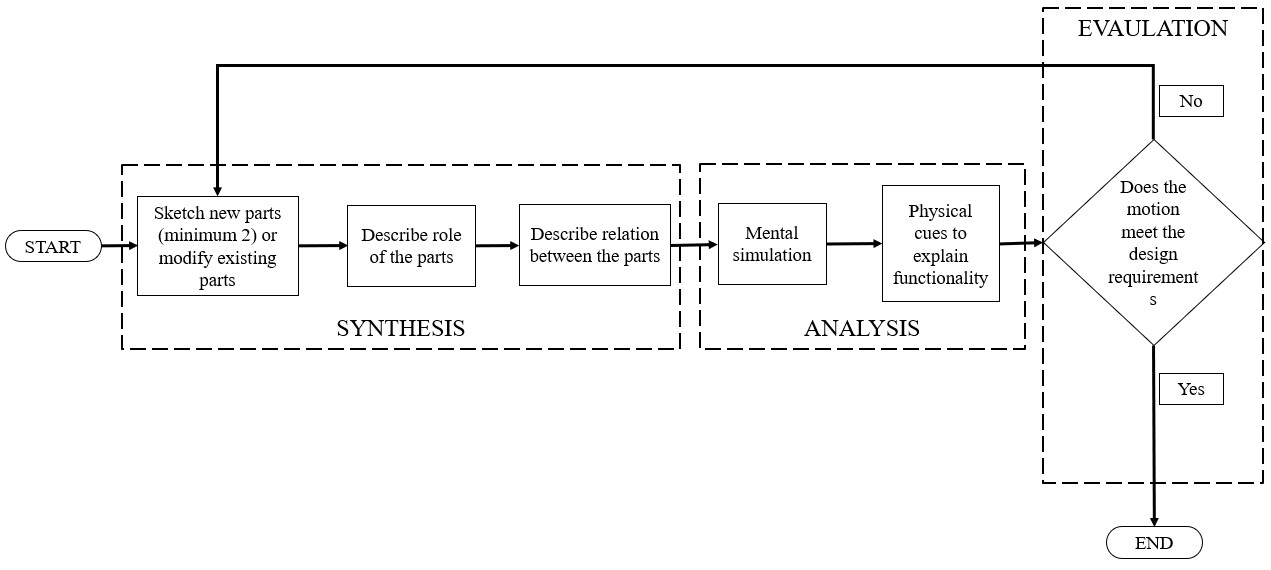}
    \caption{Methodology followed by the designers while motion exploration in \cite{ramana2020designers}}
    \label{flow_chart}
\end{figure}

Simple motions, such as rotation and translation (e.g., the base and hammer motion in the stapler, Fig. \ref{fig:common_articulations}(a)), are easy to visualize mentally and validate, whereas complex motions involving non-rotational and non-translational components (e.g., the motion of a point on the coupler link 6 in Fig. \ref{fig:common_articulations}(b)) are more challenging. The situation gets complicated as the number of components increases. From a design cognition perspective, this process of visualizing complex motion necessitates higher working memory, i.e. the designers have to mentally hold multiple components in it and ascertain relative motion \cite{babbage1826xviii, Deng2002, sweller2011cognitive}. Consequently, concept evaluation depends solely on the designers' ability to accurately visualize component movement through what \cite{Hegarty2004, Bilda2007} describe as \emph{mental simulation}. To clarify functionality, designers employ various design actions such as verbal explanations, hand gestures, and textual annotations, as shown in Fig \ref{motion_exploration_requirements}. For instance, they may represent connections through small circles for revolute joints, and arrows may be used to indicate motion types (Fig. \ref{motion_exploration_requirements}(a)). Unique gestures further illustrate concepts like spring behaviour through specific hand movements. Additionally, designers create auxiliary sketches (Fig. \ref{motion_exploration_requirements}(b)) alongside the main sketch when complexity increases due to numerous components. These auxiliary sketches help decentralize mechanisms and enhance clarity by simplifying complex interactions. 
As there is a gap between intention and realization of functionality of product concepts mentioned by \cite{GERO2004FBS}, it was concluded by study in \cite{ramana2020designers} that they often fail to meet expected performance. It is essential to verify the behaviour of the concept represented for a valid product proposal. To address this, designers must engage in an iterative synthesis, analysis, simulation, and motion validation process, which helps refine concepts into workable solutions (as shown in Fig. \ref{flow_chart}). This process is known as \textit{motion exploration}. 

Also, the form of the sub-components is not formally decided at the conceptual stage. Conventional Computer-Aided-Design (CAD) software uses primitive geometries (such as lines, triangles, etc.) for modelling. Sketches are vague and unclear \cite{Goel1991, KAVAKLI1998485}, but the geometry and topology of CAD elements are well-defined. The process of virtual prototyping and simulating the motion is resource-intensive. Thus, reducing the time and effort spent validating the concepts will significantly improve the efficiency of the product development cycle. For this, \cite{Onkar2010} reports that a sketching environment should facilitate freestyle stroke generation and concept validation through the phases of perception, cognition, and simulation that reduce the complexity of the cognitive process in sketch understanding.  
\clearpage
The present work aims
\begin{itemize}
    \item To support motion exploration in an integrated sketching environment where the main activity is concept sketching, and the designer is given feedback on the motion without deviating much from the sketching activity.
    \item To evaluate the sketching environment in terms of subjective satisfaction and mental effort. 
\end{itemize}

\section{Literature Review}
This section broadly explores three domains that affect sketching behaviour in the digital environment for articulated product concepts. 

\subsection{Sketching and Modeling}
Tversky \cite{Tversky2002} argues that, in design, sketching reflects the underlying conceptual structure of the domain. Thus, it is necessary to support the early stages of design. Towards this, computer-based techniques quickly visualise the shape created by the designers using sketching. For example, Teddy \cite{Igarashi1997} is an application that generates 3D shapes based on sketch strokes. IloveSketch, \cite{Bae2008}, an application that allows designers to create sketch strokes in arbitrarily oriented planes. Similarly, other applications help visualize the shape information. In addition to shape information, annotations represent usage and behaviour. In contrast to the shape representation in sketches, which is directly observable, the behaviour needs to be simulated and verified. An example of such an application is A Shrewd Sketch Interpretation and Simulation Tool (ASSIST) \cite{Alvarado2001}, where components are represented through line diagrams and geometric shapes are recognised from the sketches. Their dynamic behaviour is simulated, relying on gestural inputs.  Another tool is a sketching interface for finite element analysis (FEA) \cite{FEAsyJournal2017}, which extracts details such as geometry, force interactions, and boundary conditions from sketch strokes, along with the specified material model, to simulate strength behavior. Researchers \cite{wetzel2009automated,onkar2013behaviour} also emphasized the importance of visualizing motion behaviour in concept sketches. Applications like Mechanix \cite{Atilola2011} solve free-body diagrams for truss problems using a sketching interface.  An automated approach for generating ``how things work" from 3D CAD models was shown in \cite{Mitra2010} by combining shape analysis techniques with visualization algorithms. Algorithms to recognize kinematic chains of hand-drawn sketches based on training on text graphics \cite{eicholtz_kara} and neural networks \cite{nurizada2024transforming} respectively. 

\subsection{Mechanism Synthesis and Analysis}
Chase et al. \cite{chase2013} review software tools designed for the motion generation of rigid bodies. A web-based mechanism design tool, shown in \cite{Cheng2005}, can compute and visualize the kinematic parameters of the individual links of mechanisms such as the four-bar, crank-slider, geared-five-bar, six-bar linkages, and cam-follower systems. A modeller is associated with the design system in \cite{KaustubhMcCarthy2015} to synthesize six-bar and spherical mechanisms.  \cite{PurwarMotion_Gen} developed an Android and iOS application for motion generation. MechDesigner \cite{mechDesigner} is a software application specialized in designing and analyzing cams, motions, and mechanisms are achievable only through a constraint-based sketching editor. M.Sketch \cite{m_sketch_2018_journal} is developed to use technical non-experts for analyzing mechanisms. FoldMecha \cite{foldmecha_2017} is a computer-aided mechanism design system that supports the construction of mechanical papercraft. It enables users to design motion by varying component parameters. Then, the physical prototypes are built using the system-generated parts and folding nets. MechPerfboard \cite{perfboard_2018} used augmented reality (AR) for designing mechanisms for path generation problems. Haptic feedback was added by \cite{haplinkage_2020} as an extra feature to simulate hand tools like wrenches, scissors, syringes, etc. A workbench for interactive kinetic art (WIKA), \cite{wika_2020}, blends artistic exploration, mechanism building and programming. \cite{deyiIntegratedFramework2021} made an integrated approach for modelling, simulation, and optimization in the design of complex mechanical products. \cite{perfboard_2018, urbina2022representation} have used AR for mechanism design tools in academic environments.

\subsection{Usability of sketch-based mechanism design applications}
Usability studies \cite{usbility_jakob} for any system are a prime requirement for its acceptability. The system's acceptance is based on social and practical aspects. Practical aspects include usefulness, cost, reliability, etc. Usefulness is divided further into two categories viz: usability and utility. The utility is the question of whether the system is functioning as intended. Moreover, usability is `\textit{how well}' this functionality is. The usability of a system is defined in terms of specific components, viz., learnability, efficiency, memorability, error, and satisfaction. \cite{usability_measure_2006} has defined and classified usability measures into effectiveness, efficiency, and satisfaction. Usability definitions from \cite{shackel_2009} and \cite{iso2018ergonomics} do not deviate much from that given in \cite{usbility_jakob}. Evaluating a system using the above five dimensions is considered usability testing. Testing is done to determine whether the system has a direct impact on the users or not. Even though the authors of the usability studies may not use the five components for their evaluation, they may define the components that match semantically.

Digital sketching interfaces have been developed considering the limitations of the traditional sketching medium. Once they have been developed, they need to be evaluated with some criteria for usability. In  \cite{samavati2011sketch}, an overview of the important sketching interfaces is presented. \cite{EverybodyLovesSketch_2009, sketchwithhands_2016, dreamsketch_2018} used qualitative evaluation methods for usability. \cite{teddy_1999, drawingonair_2007, EverybodyLovesSketch_2009,arora_2017, sweepcanvas_2017, kinesthetic2019journal,mobi3Dsketch_2019} devised their own parameters of usability evaluation. There have been attempts to design spherical \cite{furlong1999} and spatial mechanisms \cite{Kihonge2002} in virtual reality (VR). A study on the effectiveness of traditional and VR interfaces in spherical mechanism design \cite{evans1999} proved that the former was chosen over the latter. Of course, the experiments performed had their limitations.

\subsection{Summary of Literature Review}
The literature, \cite{wetzel2009automated, dong2023towards, rizzuti2021interactive}, reveals difficulty in communicating the functionality through articulated product concept sketches. Further, sophisticated conceptual design tools for motion exploration are scarce, and conventional software uses primitive geometries to represent the structure of the sub-components in their design. \cite{ball2019advancing} also suggests an uncertainty or discrepancy between structurally- and functionally-oriented sketching, which aims to reflect the connectivity between elements and their integrated behaviour. Hence, it is important to support designers in exploring motion in articulated product concepts. It is also observed in the \cite{nikulin2019nasa, ramana2020designers} that the designers experience higher cognitive load in the conceptual stage because of the intense design activity of iterations to generate, analyse and validate the concepts (as shown in Fig. \ref{flow_chart}). Hence, verifying whether the prescribed support reduces mental effort is also essential. Overall, the main research gap that is covered in this study is the development of the concept generation tool and its evaluation in terms of subjective satisfaction and mental effort. 

\section{Methodology}
\label{tool_dev}

\begin{figure}[!h]
    \centering
    \includegraphics[width=0.8\linewidth]{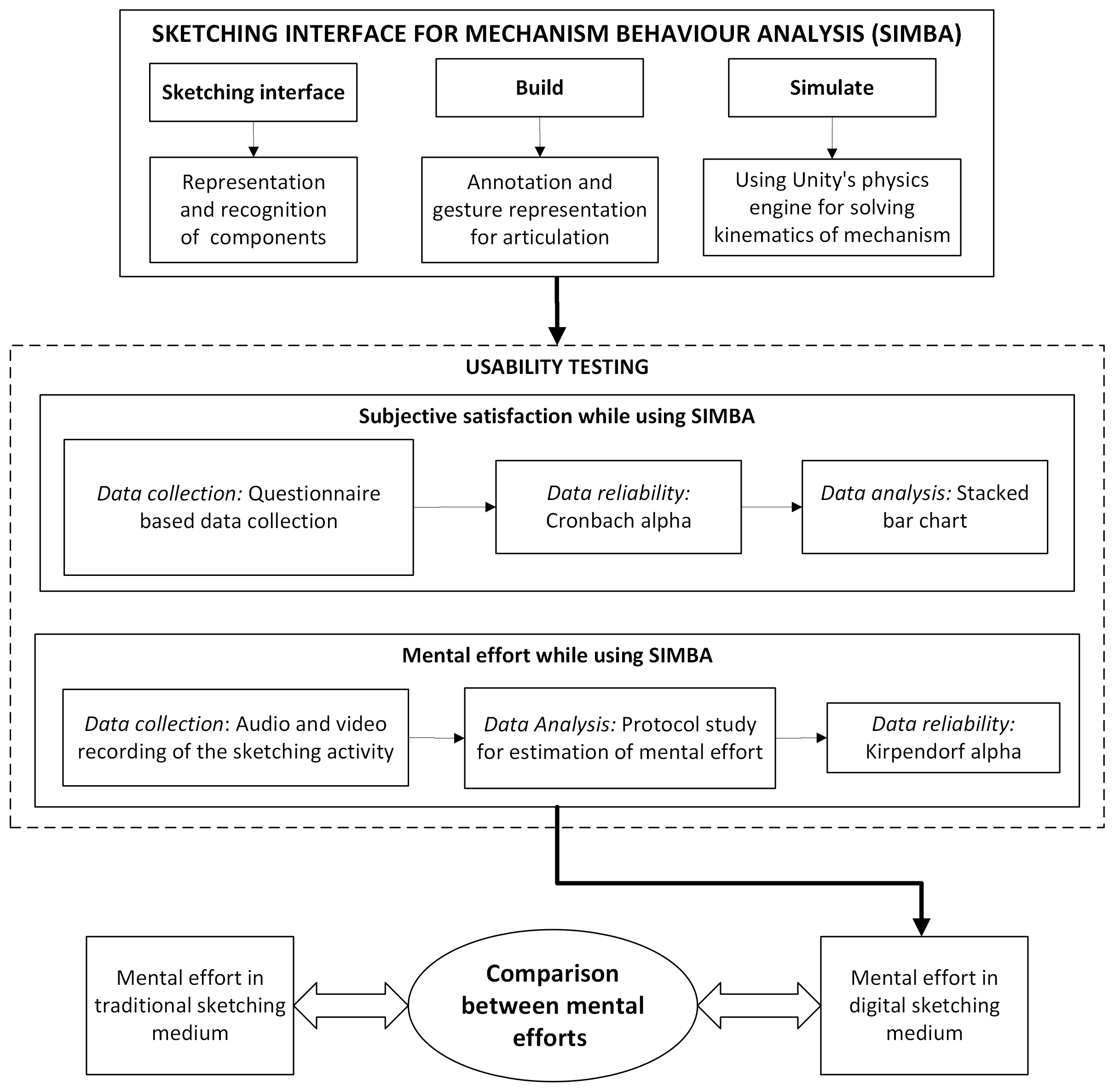}
    \caption{Methodology (Source: authors)}
    \label{fig:methodology}
\end{figure}
To investigate whether a support system will help the designer with motion exploration at the conceptual design stage, a 2-step development and evaluation methodology (Fig. \ref{fig:methodology}) is followed. Unity$^{TM}$ platform is a game-engine where one can place game objects in a scene and program their interactions between them. Sketching Interface for Mechanism Behaviour Analysis (SIMBA) was developed using Unity$^{TM}$. During this application development, the gestures and annotations of the traditional sketching medium (as mentioned in the introduction section) are characterised digitally. Sketch recognition algorithms are used from \cite{Onkar2010} for the representation and recognition of components. The kinematic relations between the components are characterised based on the gestures and annotations, \cite{onkar2013behaviour}, made by the designers. Unity's physics engine is used to simulate the desired motion.

Using quantitative methods, usability testing of SIMBA is performed to estimate the affective and cognitive responses of the participants. The design students were selected as participants since they are novices in motion exploration, which helps identify different use cases compared to a professional designer. A design experiment was conducted where the students were asked to generate concepts for a given design problem using SIMBA. Firstly, the affective response is measured using subjective satisfaction parameters \cite{hornbaek2006current} viz., ease of use, want-to-use, fun-to-use, and intuitiveness.  Secondly, the cognitive response is measured by mental effort estimation. Audio-video recordings of design experiments are analysed using the protocol study \cite{ramana2020designers} to estimate the mental effort involved in motion exploration while sketching. The collected data are tested for reliability in both cases. The results obtained are compared with those of a traditional sketching environment. Bootstrap algorithms in IBM SPSS \cite{SPSS}, are used to test the significance of the results statistically.

\section{Tool development}
To tackle the aforementioned challenges, SIMBA was created using Unity version 2021.3.7f1 on a Lenovo ThinkStation equipped with an Intel(R) Xeon(R) Gold 5118 CPU running at 2.30GHz (48 CPUs) and 64 GB of RAM. This ThinkStation also featured an NVIDIA Quadro P6000 graphics card. Sketching was facilitated through a haptic device from 3D Systems, specifically the Geomagic Touch. The software application is designed for product designs involving multiple components that move relative to one another in parallel planes, requiring that the product sketches incorporate planar mechanisms. The tool was developed using the Unity and Visual Studio game engines, with sketching performed via the Geomagic Touch haptic device.

\subsection{Characterisation of design actions for motion exploration for SIMBA}
Using the observations made on the traditional sketching medium, certain features are developed in SIMBA to incorporate its nature. Each of the SIMBA features is categorised based on these observations as shown in Table \ref{tab:tradSketch_SIMBA_Mapping}. 

\begin{table}[!ht]
    \centering
    \caption{Observations in sketching activity (left column) are mapped to SIMBA features (right column) (Source: authors)}
    \label{tab:tradSketch_SIMBA_Mapping}
    \begingroup
    \small
    \setlength{\tabcolsep}{6pt} 
    \renewcommand{\arraystretch}{1.2} 
    \begin{tabular}{>{\raggedright\arraybackslash}m{60mm} >{\raggedright\arraybackslash}m{60mm}}
        \toprule
        \textbf{Observations in traditional sketches} & \textbf{SIMBA features} \\
        \midrule
        Incremental sketching of parts & Draw/sketch, Component recognition, Connections recognition, Tracing over mechanism, Attaching objects \\
        \addlinespace
        Uncertainty of the functionality and testing validation of the concept & Interactive simulation, Coupler curve generation, Joint-to-joint distance manipulation \\
        \addlinespace
        Design actions for explaining functionality & Connections recognition, simulation \\
        \addlinespace
        Auxiliary sketches & Multiple instances sketches \\
        \bottomrule
    \end{tabular}
    \endgroup
\end{table}

Using the tool, the designers can sketch the articulated product concepts. The tool can recognize links and joints and simulate planar mechanisms. The application has three tabs, namely Sketch, Build, and Simulate. Each tab has a unique colour, which indicates the active state of operation to the designer. Each tab has different features to facilitate motion exploration of the sketch, which are explained below. 
\begin{figure}
    \begin{center}
        \subfloat[Sketch features]{\includegraphics[width=0.5\textwidth,keepaspectratio]{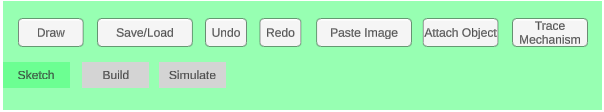}}\\
        \subfloat[Build features]{\includegraphics[width=0.3\textwidth,keepaspectratio]{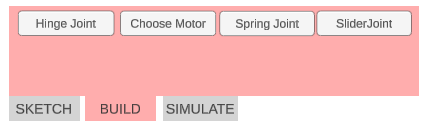}}
        \subfloat[Simulate features]{\includegraphics[width=0.3\textwidth,keepaspectratio]{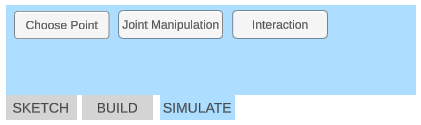}}
    \end{center}
    \caption{SIMBA features (colour of the widget indicates the active state of operation) (Source: authors)}
    \label{SIMBA_tabs}
\end{figure}

\begin{enumerate}
    \item \textit{Sketch}:
    The sketch tab (Fig. \ref{SIMBA_tabs}(a)) facilitates the creation of links and objects using the haptic stylus. Pressing the sketch button displays a message that lets the user know that sketching is enabled. Users can attach other mini sketches or images to links. The different features on the sketch tab are sketch, save, load, undo, redo, paste images, attach objects to links and trace mechanism. 
    \item \textit{Build}:
    The build tab (Fig. \ref{SIMBA_tabs}(b)) contains features that are necessary to provide motion constraints to the links that were drawn earlier. This action is performed by indicating joint constraints (rotary or slider) using sketch-based gestures. 
    \item \textit{Simulate}:
    The simulate tab (Fig. \ref{SIMBA_tabs}(c)) contains features to simulate the mechanism and visualize the motion of different links. This also includes point trajectories, joint manipulation, and interactive simulation.
\end{enumerate}

\subsection{SIMBA features}    
The SIMBA features are individually explained in this section in the order of their workflow for simulating articulated product concepts. 
\begin{enumerate}
   \item {\textbf{\textit{Recognition of Links/Components/Rigid bodies}}}: The form of an object depicted in a sketch is less explicit compared to a physical prototype or computer model, relying heavily on the designer's interpretation which is significantly shaped by Gestalt principles of perception and the skill involved in creating the sketch. Gestalt laws of the perceptual organization \cite{Wertheimer1923} help in perceiving the composition.
    \begin{figure}
        \begin{center}
            \subfloat[]{\includegraphics[width=0.3\textwidth,keepaspectratio]{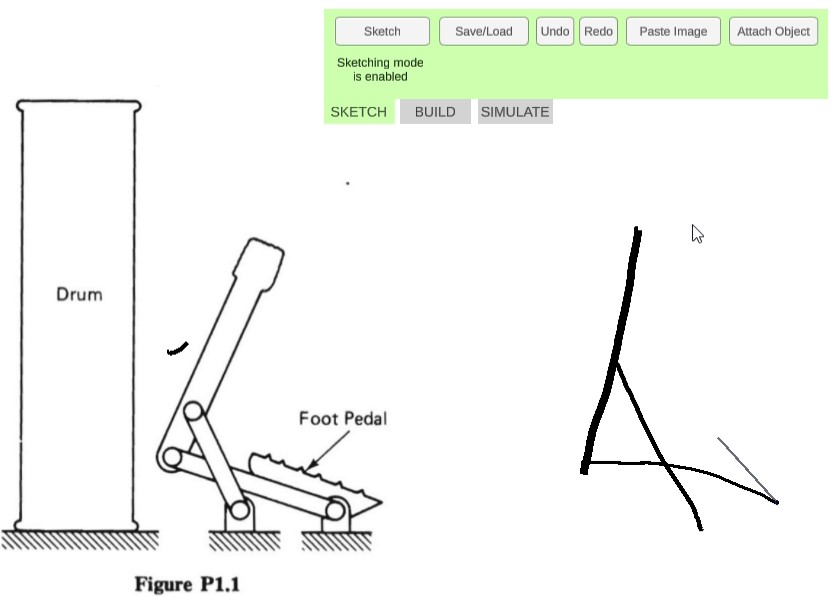}}
            \subfloat[]{\includegraphics[width=0.3\textwidth,keepaspectratio]{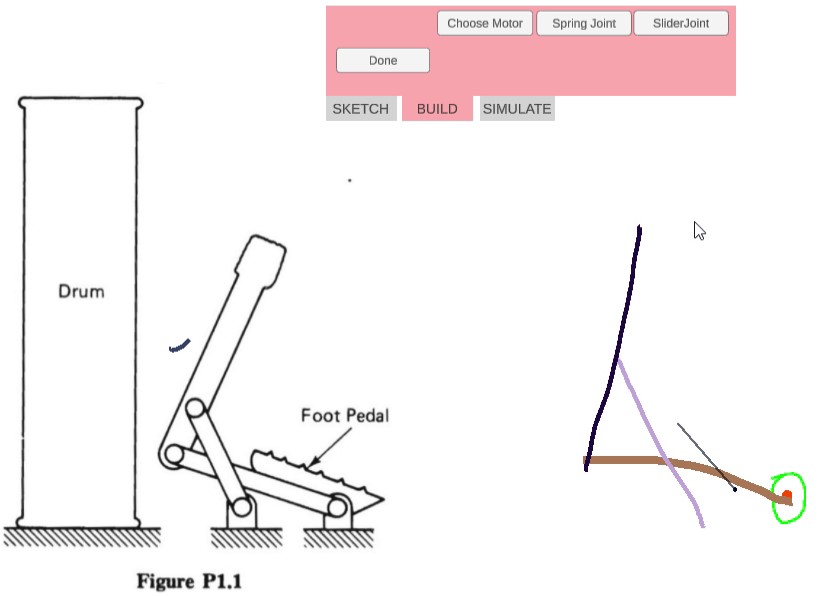}}
            \subfloat[]{\includegraphics[width=0.3\textwidth,keepaspectratio]{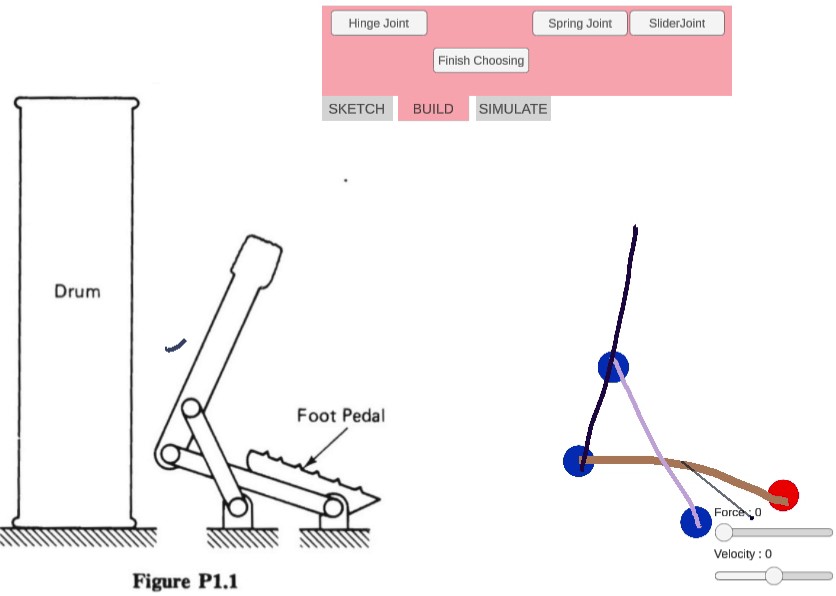}}\\
            \subfloat[]{\includegraphics[width=0.3\textwidth,keepaspectratio]{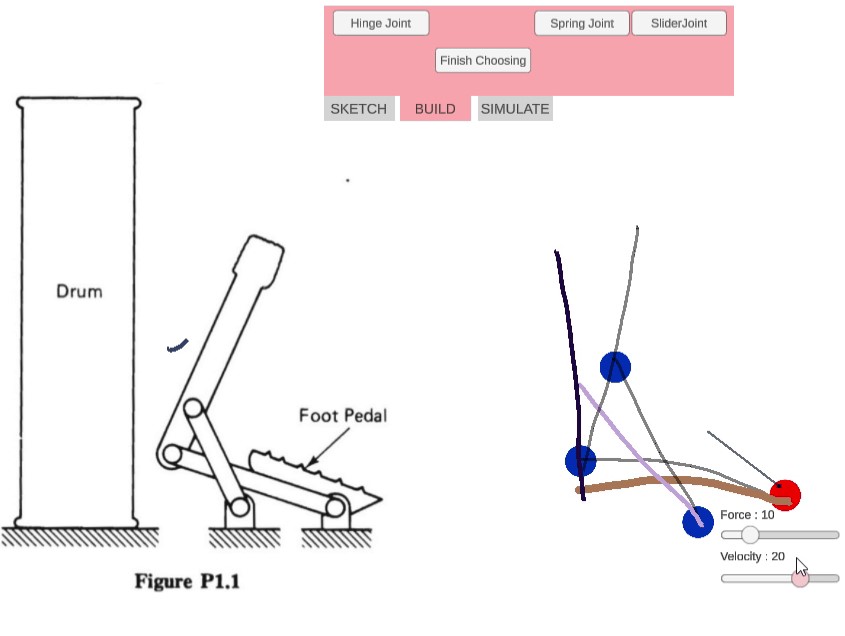}}
            \subfloat[]{\includegraphics[width=0.3\textwidth,keepaspectratio]{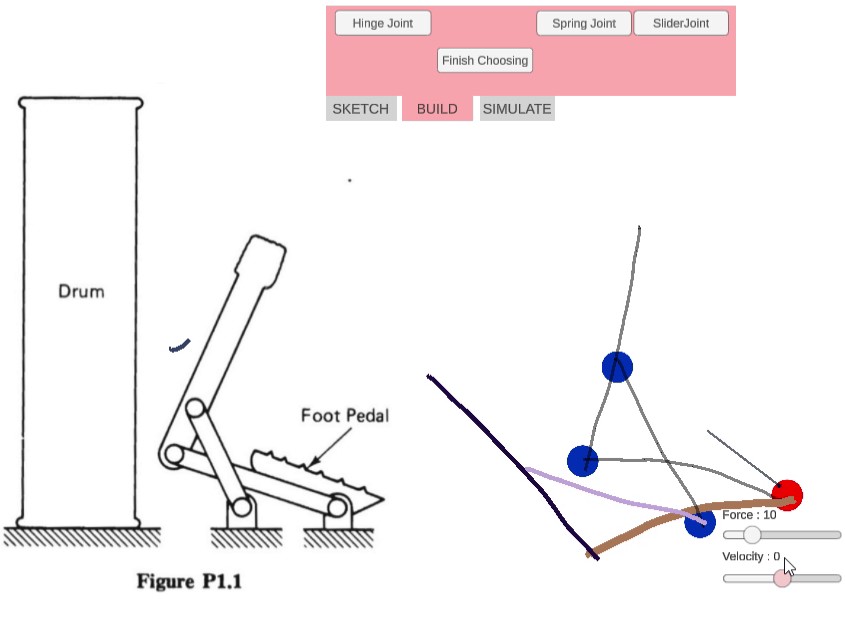}}
            \subfloat[]{\includegraphics[width=0.3\textwidth,keepaspectratio]{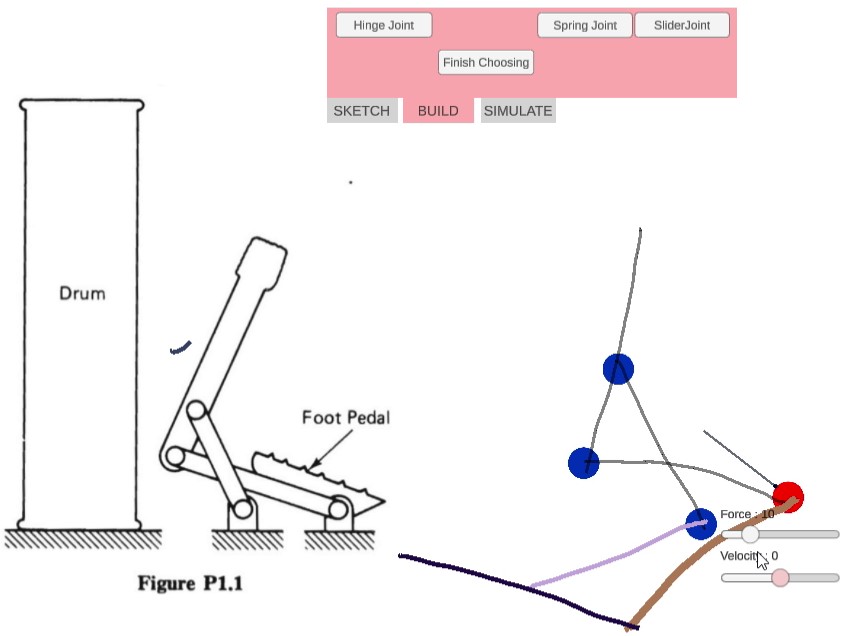}}\\
            \subfloat[]{\includegraphics[width=0.3\textwidth,keepaspectratio]{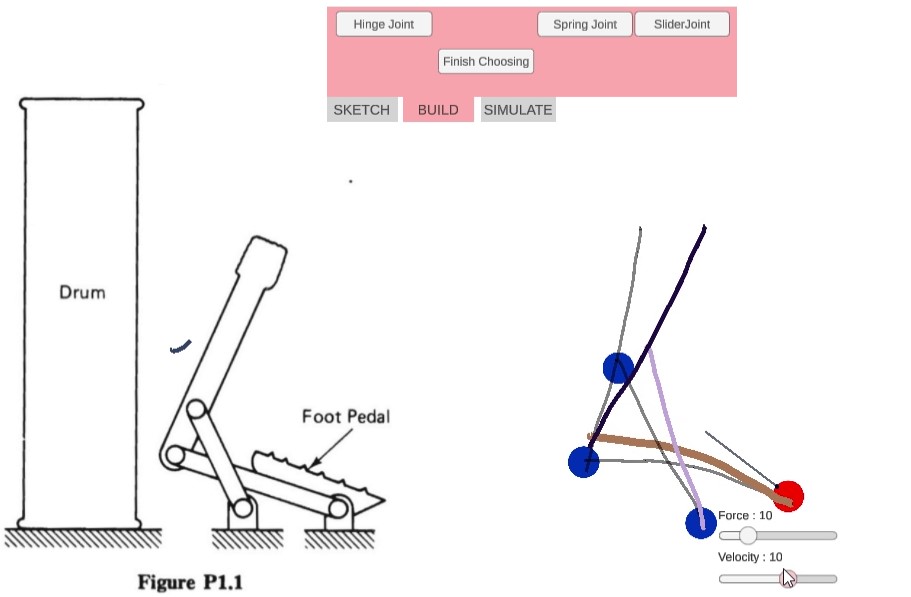}}
            \subfloat[]{\includegraphics[width=0.3\textwidth,keepaspectratio]{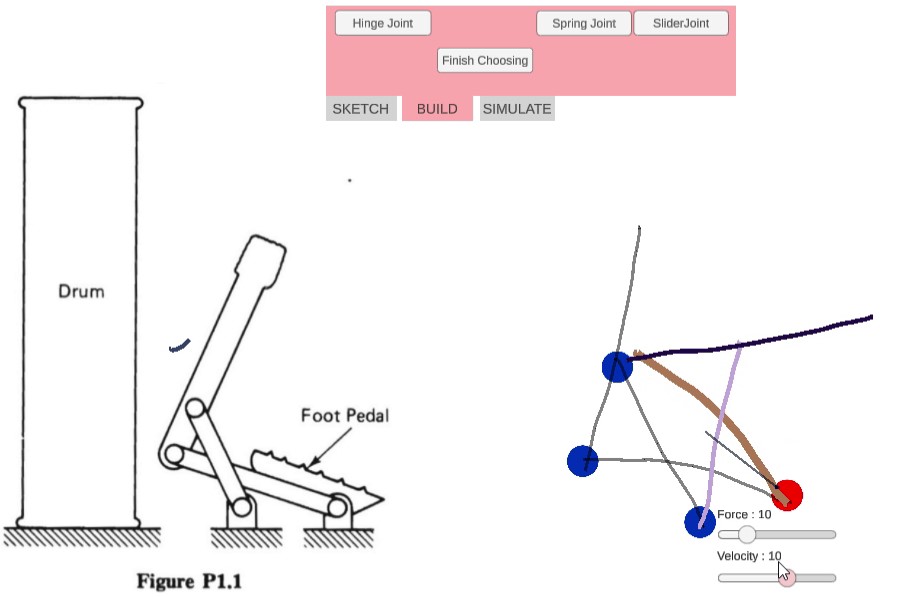}}
            \subfloat[]{\includegraphics[width=0.3\textwidth,keepaspectratio]{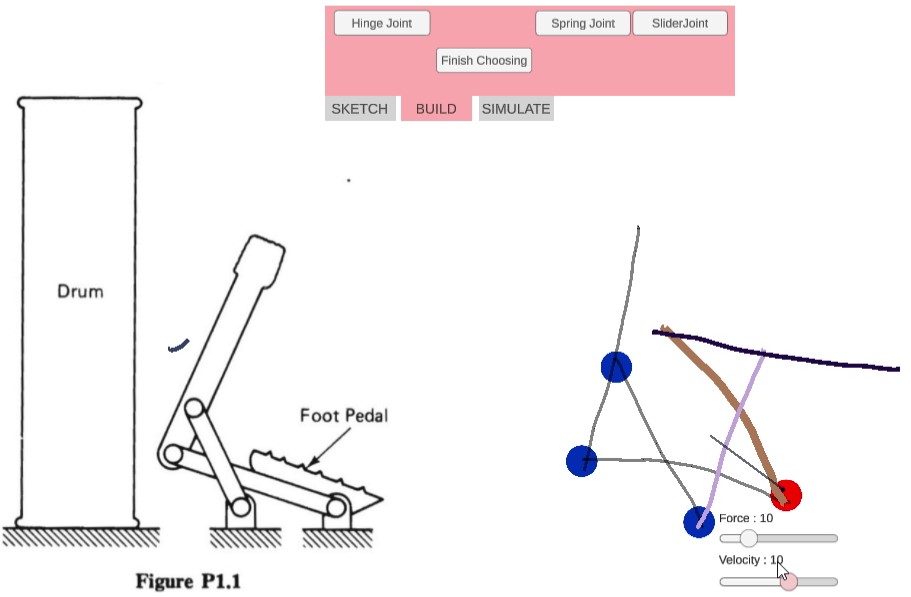}}
        \end{center}
        \caption{Stages in SIMBA (a) Sketching (b) Marking the rotary joints (c) Selecting an input joint (d)-(f) Input given in anti-clockwise direction (g)-(h)  Input given in clockwise direction. Note: Videos associated with this figure are in the supplementary material. (Source: authors) }
        \label{ex_1_simulation}
    \end{figure}
    Consider the example of the drum-beating mechanism as shown in Fig. \ref{ex_1_simulation}(a). It has four components, and the system recognises the different sub-components of the sketched concept shown in Fig.\ref{ex_1_simulation}(b). Each component is identified by a unique colour. 
        \begin{figure}
            \centering
            \subfloat[]{\includegraphics[width=0.3\textwidth,keepaspectratio]{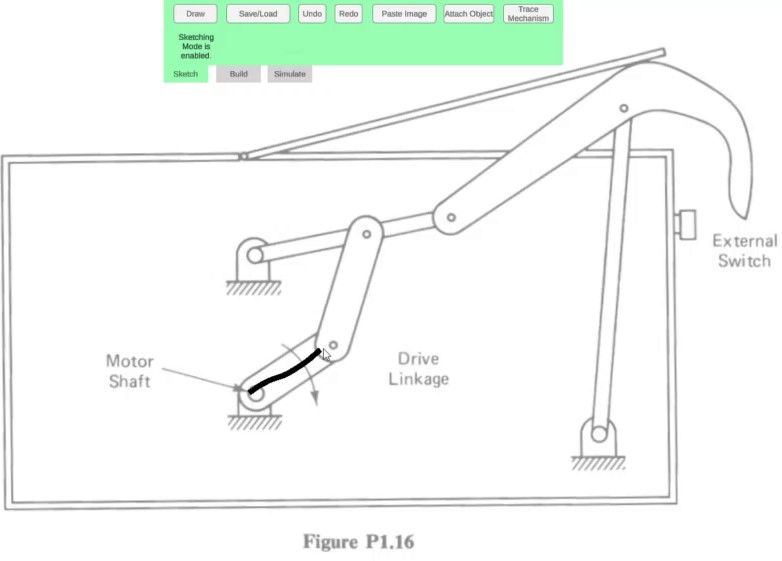}}
            \subfloat[]{\includegraphics[width=0.3\textwidth,keepaspectratio]{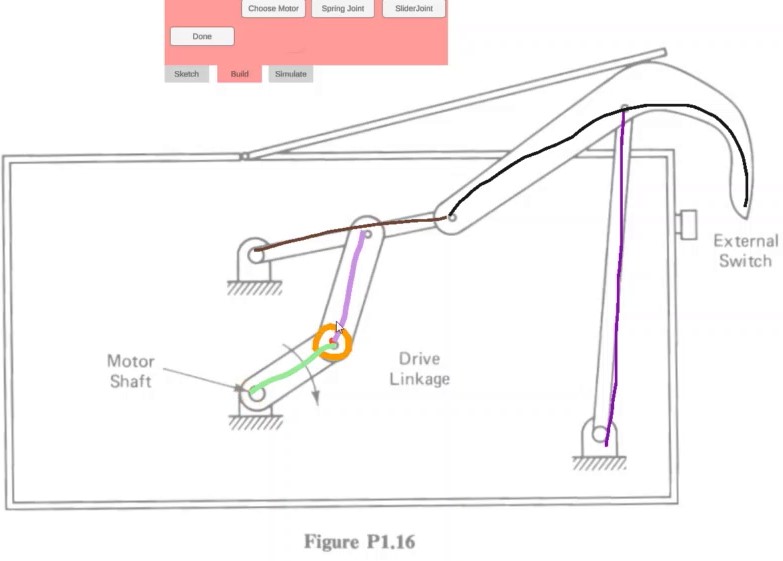}} 
            \subfloat[]{\includegraphics[width=0.3\textwidth,keepaspectratio]{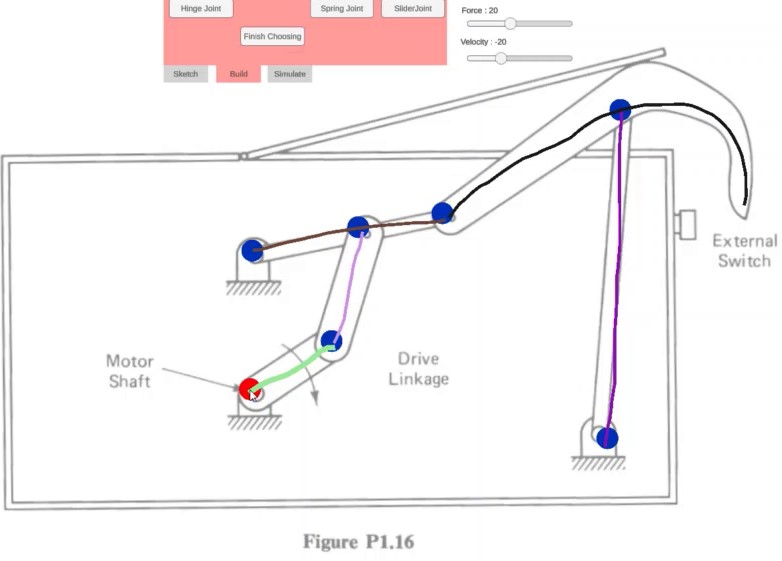}} \\
            \subfloat[]{\includegraphics[width=0.3\textwidth,keepaspectratio]{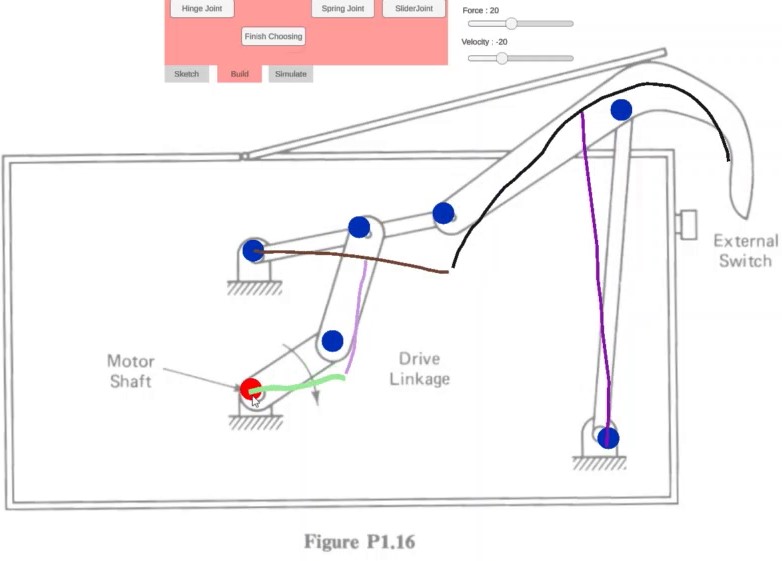}}
            \subfloat[]{\includegraphics[width=0.3\textwidth,keepaspectratio]{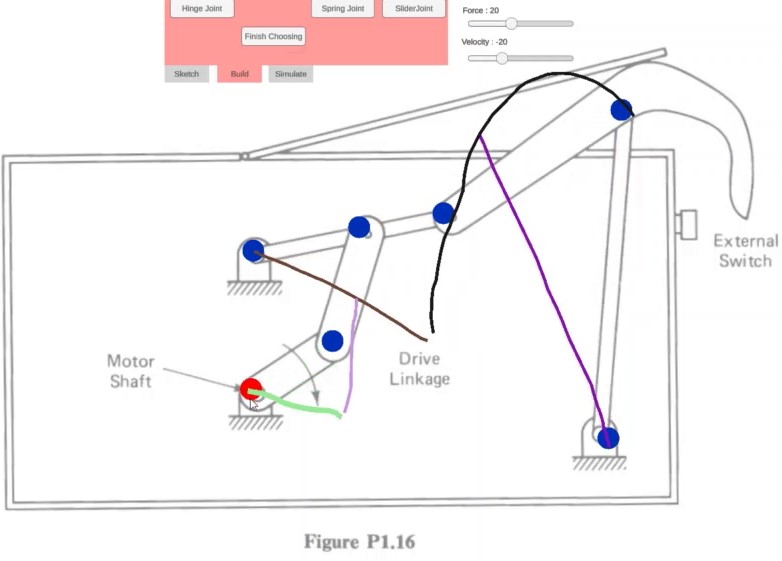}}
            \subfloat[]{\includegraphics[width=0.3\textwidth,keepaspectratio]{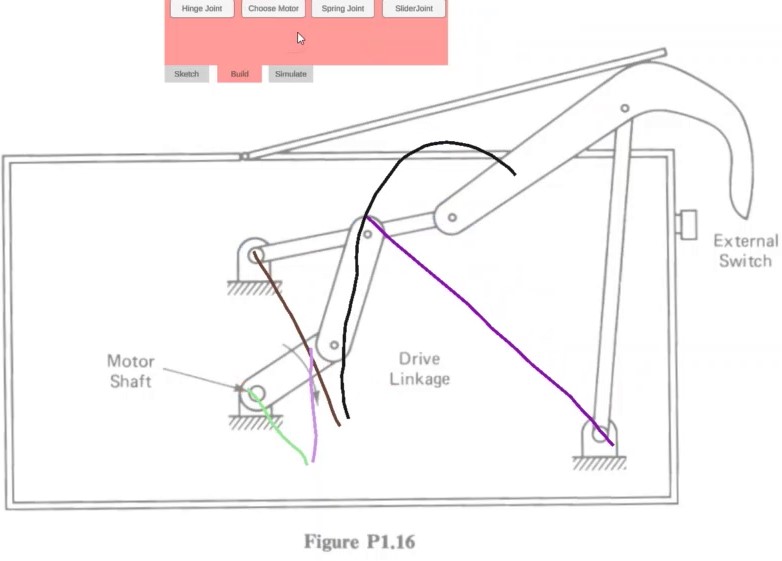}} \\
            \subfloat[]{\includegraphics[width=0.3\textwidth,keepaspectratio]{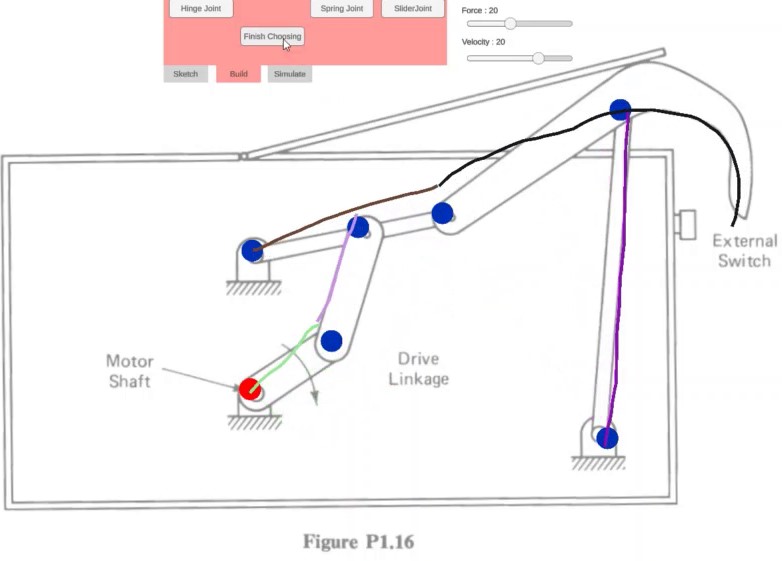}}
            \subfloat[]{\includegraphics[width=0.3\textwidth,keepaspectratio]{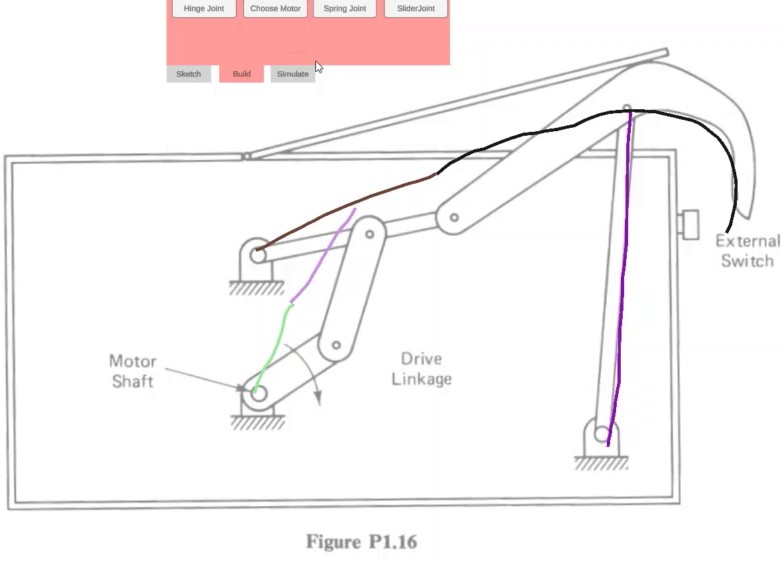}}
            \subfloat[]{\includegraphics[width=0.3\textwidth,keepaspectratio]{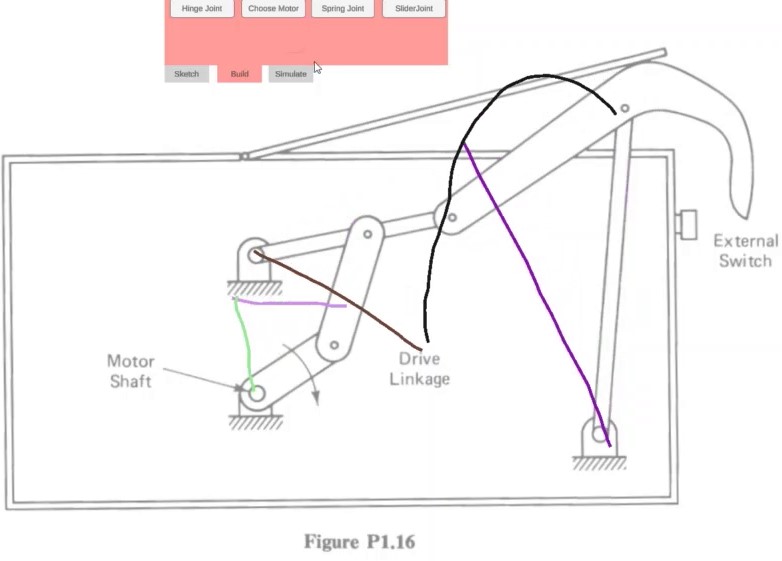}}
            \caption{ Trace mechanism feature (a) Sketching after inserting an image in the environment (b) Marking the rotary joints (c) Selecting an input joint (d)-(f) Input given in clockwise direction (g)-(h)  Input given in anti-clockwise direction}. Videos associated with this figure are in supplementary material. (Source: authors).
            \label{TracingMechanism}
        \end{figure}

    \item \textbf{\textit{Connections marking and their recognition}}: The type of joints defines the relationship between the motions of two links. In this work, only two motion constraints are considered: hinged and sliding. In hinged constraints, two components are fixed at a point and can only rotate around that point. In concept sketches, designers mark hinge locations by drawing small circles at the junction of the two links. Similarly, designers define constraints in this interface by drawing circular gesture strokes. The hinge location is determined as the average of the coordinates along the gesture stroke, as shown in Fig.\ref{ex_1_simulation}(b) and Fig.\ref{ex_1_simulation}(c).
    The mechanism is built using the connections and is ready for simulation as shown in \ref{ex_1_simulation}(c). The connections are shown as solid blue circles. The input for the mechanism is given at the location. Once the system recognises the input, the blue circle turns red. 
    
    In a sliding type of constraint, one component moves relative to the other along a predefined line of motion. This helps the designers simulate concepts with linear component motion concerning other components. This type of constraint is defined by a line along which a component translates with respect to other components. First, the designer has to select the two components between which the constraint has to be defined, and then the designer has to select the components. Then, draw a constraint line stroke to determine the sliding direction. 

    An intuitive way to describe motion between two links can be used to choose rotary constraints. Here, to constrain the rotary motion between two links, a circle or arc is drawn by pressing the second button of the stylus such that the circular arc intersects the two links. This helps the system to recognise the rotary joint between the two links. A straight line is drawn to intersect the two links for a slider joint, and the direction of motion is along it. Similarly, spring anchor points are chosen. The input is given to the joint, where the user will specify non-zero values for force and velocity parameters using the slider UI feature.

    \item {\textbf{\textit{Tracing over the mechanism} \cite{Kalyan_Sumedh_Onkar2023}}}: Sometimes, the designers want to evaluate existing product concepts. This feature allows the designers to trace over an image and simulate the existing mechanism. The system will identify the components and their connections to help the user explore the motion. Fig. \ref{TracingMechanism} shows one of the tracings over the mechanism. 
        
     \item \textbf{\textit{Interactive simulation}}: In physical prototypes, motion exploration is informal; the designer manually manipulates the chosen component. In contrast, computer models offer a structured approach with predefined fixed links, input pairs, and input functions. This allows the designer to select any input pair and examine the motion of the connected components, closely mimicking the experience of manipulating a physical prototype. This setup makes it easier to assess any functional uncertainties.

     \item \textbf{\textit{Coupler curve generation:}} Most of the time, design requirements are represented as motion requirements. These requirements include the path traced by a point on a rigid body or link, the velocity of the point, the dwell time during the motion, etc. All these facilities can be visualised in the current sketching environment in the concept generation, thereby helping designers to judge whether the mechanism meets the motion requirement. This is done by selecting a point of interest $P$ on any link, and the software traces the path of the point. (see Fig. \ref{ex_2_simulation}(a)). 

        \item \textbf{\textit{Joint location manipulation}}:
        \begin{figure}
            \begin{center}
                \subfloat[Path tracing]{\includegraphics[width=0.4\textwidth,keepaspectratio]{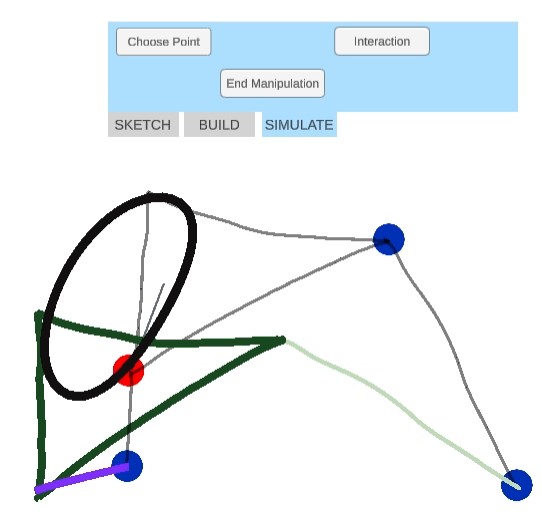}} \vspace{0.5cm}
                \subfloat[Path of the point changes after joint manipulation]{\includegraphics[width=0.345\textwidth,keepaspectratio]{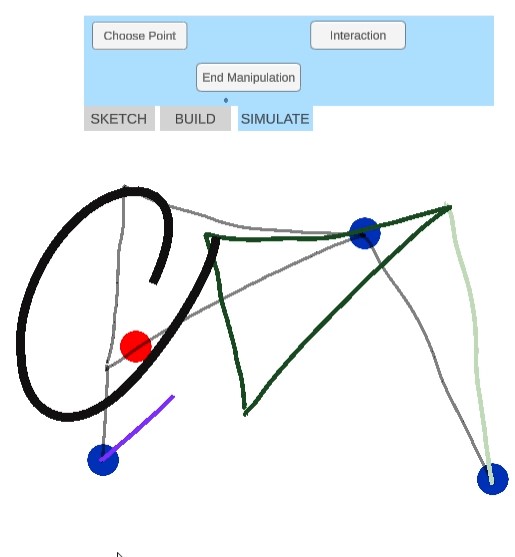}}
            \end{center}
            \caption{Coupler curve generation and joint-to-joint distance manipulation. (Source: authors).}
            \label{ex_2_simulation}
        \end{figure}
        Designers sometimes make multiple instances of the exact mechanisms with slight link length modifications that utilise space and time. Of course, older designs also need to be reviewed. To overcome the problem of wasting time and space in paper-based mode sketching, the sketching environment is facilitated with an option to manipulate the link's length. First, the joint (red in Fig. \ref{ex_2_simulation}) is selected, which will be changed. This is done by selecting two adjacent links. This implies that the links associated with the strokes are selected. After this, the link (Figure \ref{ex_2_simulation} (b)) whose length will be changed is selected. Then, the joint is moved in the X and Y direction of the sketch plane to a desired location. In terms of mechanism definition, this implies a change in link lengths. The change in link dimensions can be observed in the change in shape of the coupler curve (black). Two instances of mechanisms can be drawn as shown in Fig. \ref{multiple_mechanisms}.

    \item \textbf{\textit{Multiple instances of mechanisms}}:
   The designers drew many auxiliary sketches besides the main sketch, where the overall functionality exists. The main sketch had the main functionality, but there were certain instances where the number of components had become too large, which was difficult for the designer to handle. So, the designers used to draw auxiliary sketches on one side apart from the main sketch. This helped them decentralise the mechanism to get clarity as the complexity decreased. It is observed that designers drew auxiliary sketches to explore and then emphasise the accepted parts by drawing darker strokes. Sometimes, the same concept needs to be redrawn with different dimensions, which occupy space on the paper, as shown in the Figure. \ref{motion_exploration_requirements}(b).
   
    \begin{figure}
        \centering
        \includegraphics[width=0.8\linewidth] {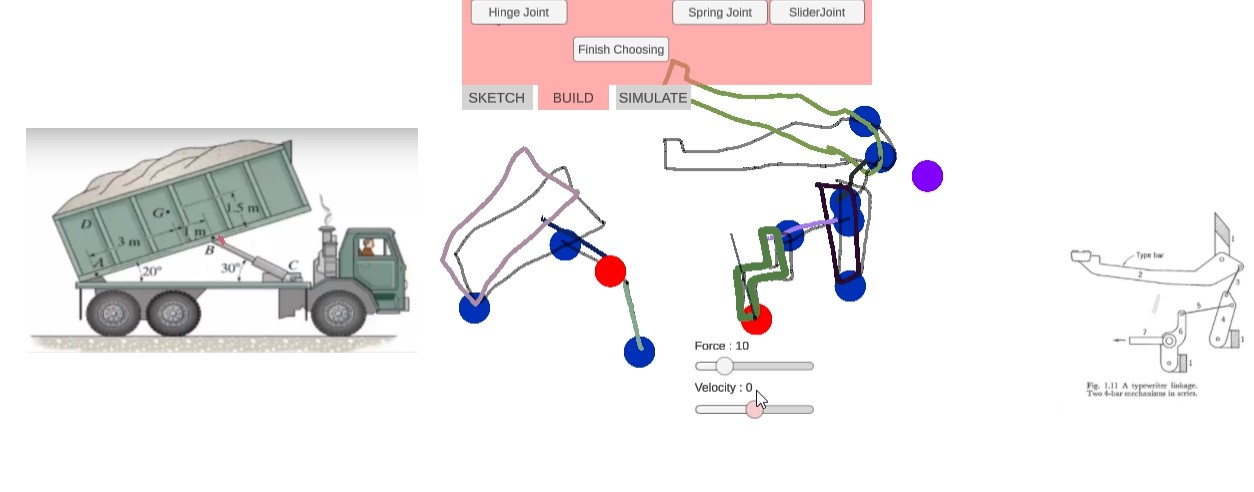}
        \caption{Two mechanisms can be drawn in the same sketching environment, images are taken from \cite{erdam1998mechanism}. Videos associated with this figure are in supplementary material. (Source: authors)}
        \label{multiple_mechanisms}
    \end{figure}
    

\end{enumerate}

\begin{figure}
    \begin{center}
        \subfloat[Experimental setup]{\includegraphics[width=0.4\textwidth,keepaspectratio]{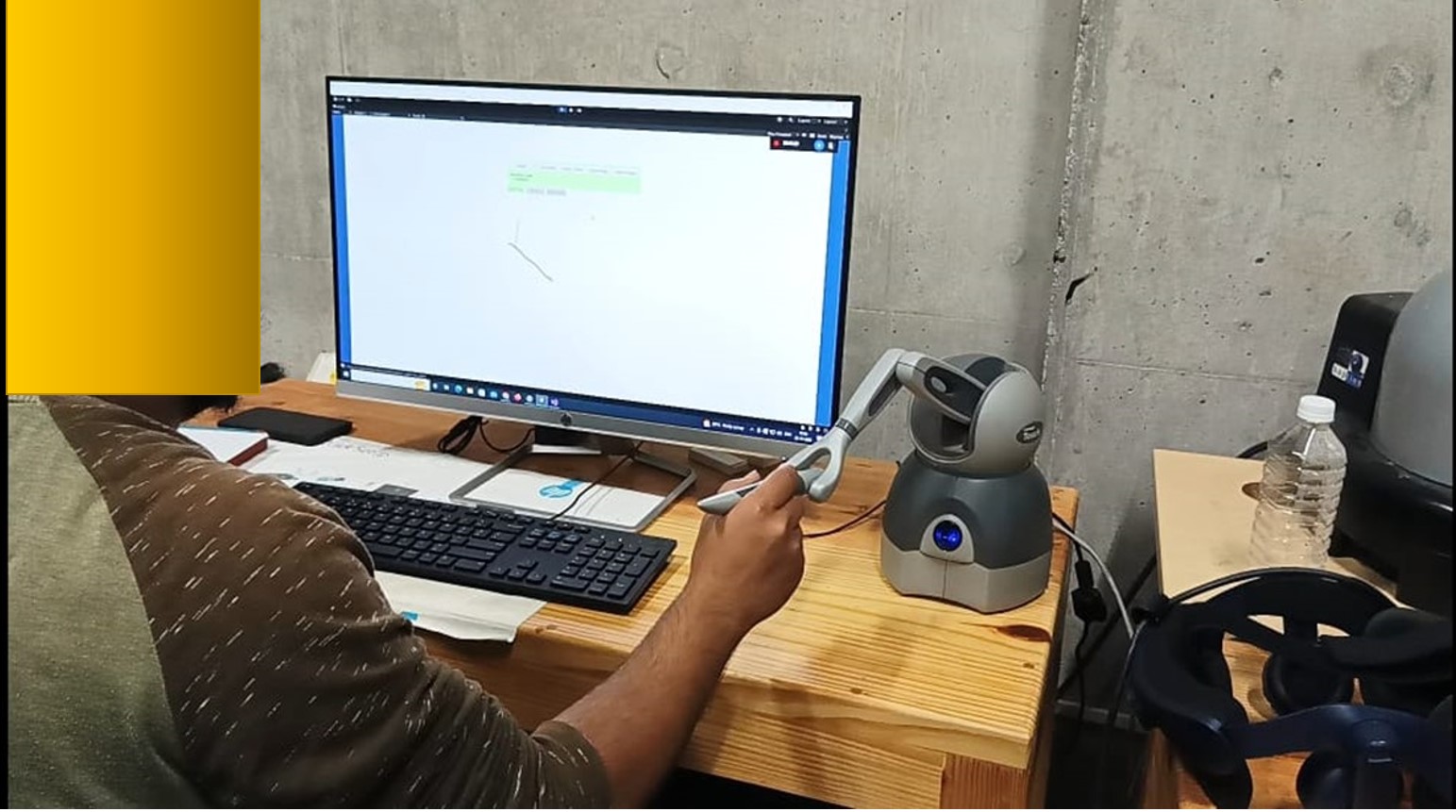}}
        \quad
        \subfloat[Processed PIP video ready for coding]{\includegraphics[width=0.4\textwidth,keepaspectratio]{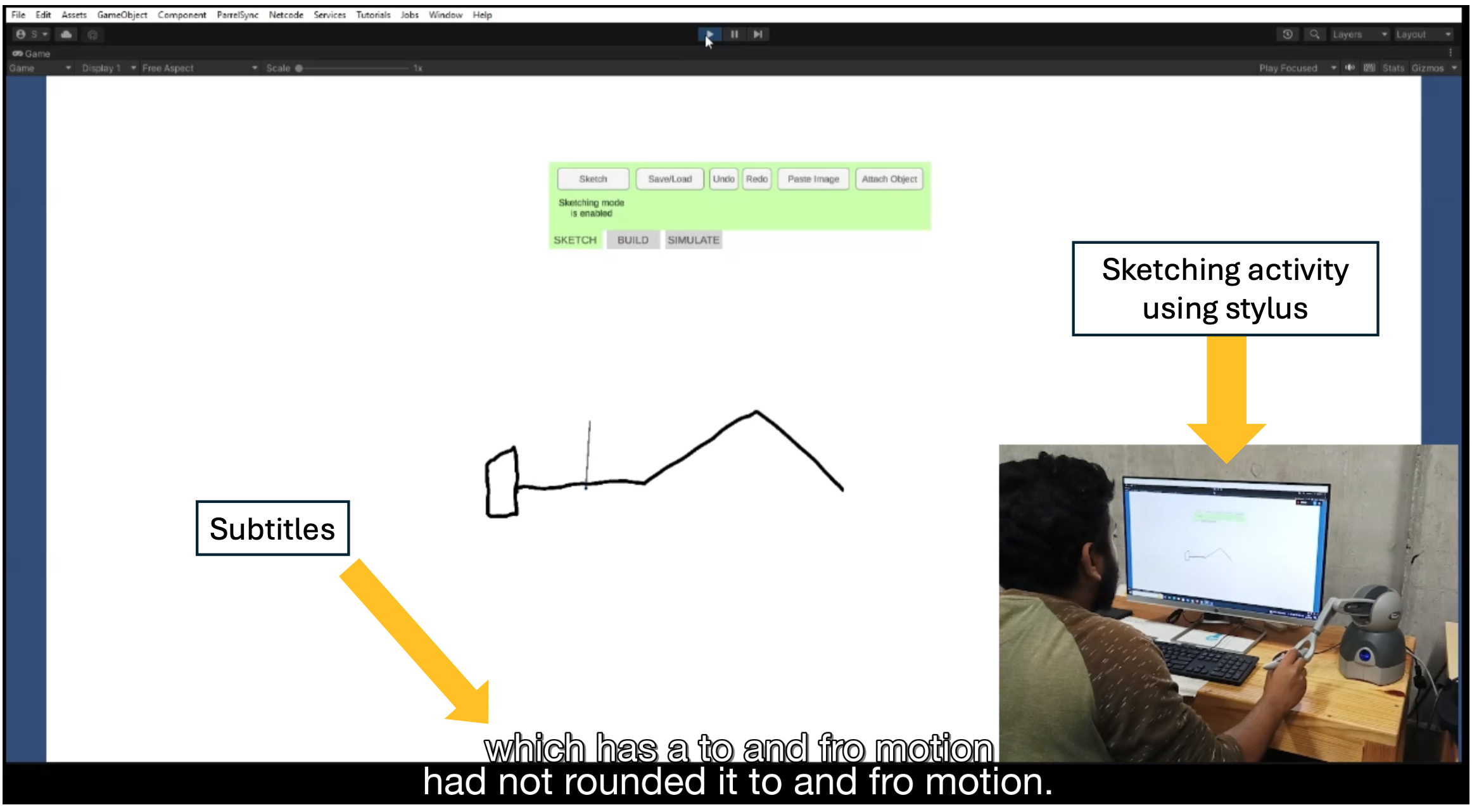}}
    \end{center}
    \caption{Experimental setup and data processing. (Source: authors).}
    \label{fig:exp_setup_AND_dataProcess}
\end{figure}

\section{Usability testing}
A design experiment is conducted to evaluate SIMBA. At first, subjective satisfaction was measured using parameters mentioned in the methodology section 3. Later, the estimation of the mental effort in motion exploration in SIMBA is shown. The comparison of mental effort in motion exploration with that of a traditional sketching environment is shown later. 
\subsection{Experiment design}
14 participants (10 males and 4 females)  were involved in this experiment. The participants were studying for Bachelor's and Master's degrees in Design (at the time of the experiment). The participants were in the age group of 18 to 32 years. Each participant experimented separately. The participants were given a design task to generate as many concepts as possible using the software application that satisfied a given design criterion. The problem statement was to \textit{"design a mechanism for an underwater submarine where the desired motion was similar to that of the hands when humans swim underwater"}. The desired motion was verbally described to the participants.  The design requirements were clearly explained, and the participant was given time to think. Any doubts from the participants were clarified. If the participant had any doubts regarding the software usage, they were to raise their hand. Other advice, especially regarding the concept, was politely discouraged, but they were encouraged to test their doubts on the software. The design activity was video recorded. The participants were to think aloud about their design actions during the experiment. The maximum time for the sketching activity was 45 minutes. In addition, the participants were asked not to worry about the correctness of the outcome of the concept. If they thought the concept would work, it had to be saved. This is to make their design activity concept-centric rather than software-based. Fig. \ref{fig:exp_setup_AND_dataProcess}(a) shows the whole experiment setup. Once the experiment was finished, the participants were asked to complete a questionnaire about their experience while using SIMBA for data collection. The data of 14 participants was used to measure the subjective satisfaction. Out of these, 4 participants' data were used to measure the mental effort involved in motion exploration while using SIMBA.  

\subsection{Subjective satisfaction of SIMBA}
The collected data has to be analysed to estimate the designers' subjective satisfaction. The test/questionnaire is also verified for reliability using Cronbach's alpha. 
\subsubsection{Data collection}
The questionnaire asks the participants to rate the features of the software application using parameters. Each question referred to a specific feature evaluated using the parameters. The following features of the software were sketching(F1), building (F2), attaching objects and images (F3), motion visualisation (F4), path tracing F5), and joint manipulation (F6). The four aspects of subjective satisfaction are easy to use (EU), fun to use (FU), want to use (desires) (WU), and intuitiveness (INT). So, WU3 refers to the measure of the want-to-use parameters for attaching objects and images. 

For each data point, the participant has their choice, viz., a five-point Likert scale of 1-5 is used where 1 represents Strongly Disagree and 5 represents Strongly Agree. Each question was asked about a specific feature of the software. After completing the questionnaire, the participants were interviewed about their overall experience and shortcomings of the software application. The answers were audio recorded. The questionnaire was filled out by the participants using a Google form. The questionnaire was set per Likert’s scale \cite{likert1932technique}. The authors have taken approval from the ethics committee for the experimentation and data collection method. There were no duplicates in the questionnaire. Each feature is evaluated against the four factors, and the results have been plotted in Fig. \ref{Results}. 

\subsubsection{Data analysis and Results}

The Cronbach's alpha \cite{cronbach1951coefficient} is estimated using the IBM SPSS software \cite{SPSS} to calculate the reliability of internal consistency among the questionnaire items. The resulting Cronbach's alpha for the current questionnaire is 0.871, well within the acceptable range of 0.7 to 0.95 \cite{devellis2012scale}. The primary aim of the study is to evaluate participants' subjective satisfaction regarding each feature based on specific parameters, rather than comparing satisfaction across different features

\begin{figure}
    \begin{center}
        \subfloat[Ease of use]{\includegraphics[width=0.4\textwidth,keepaspectratio]{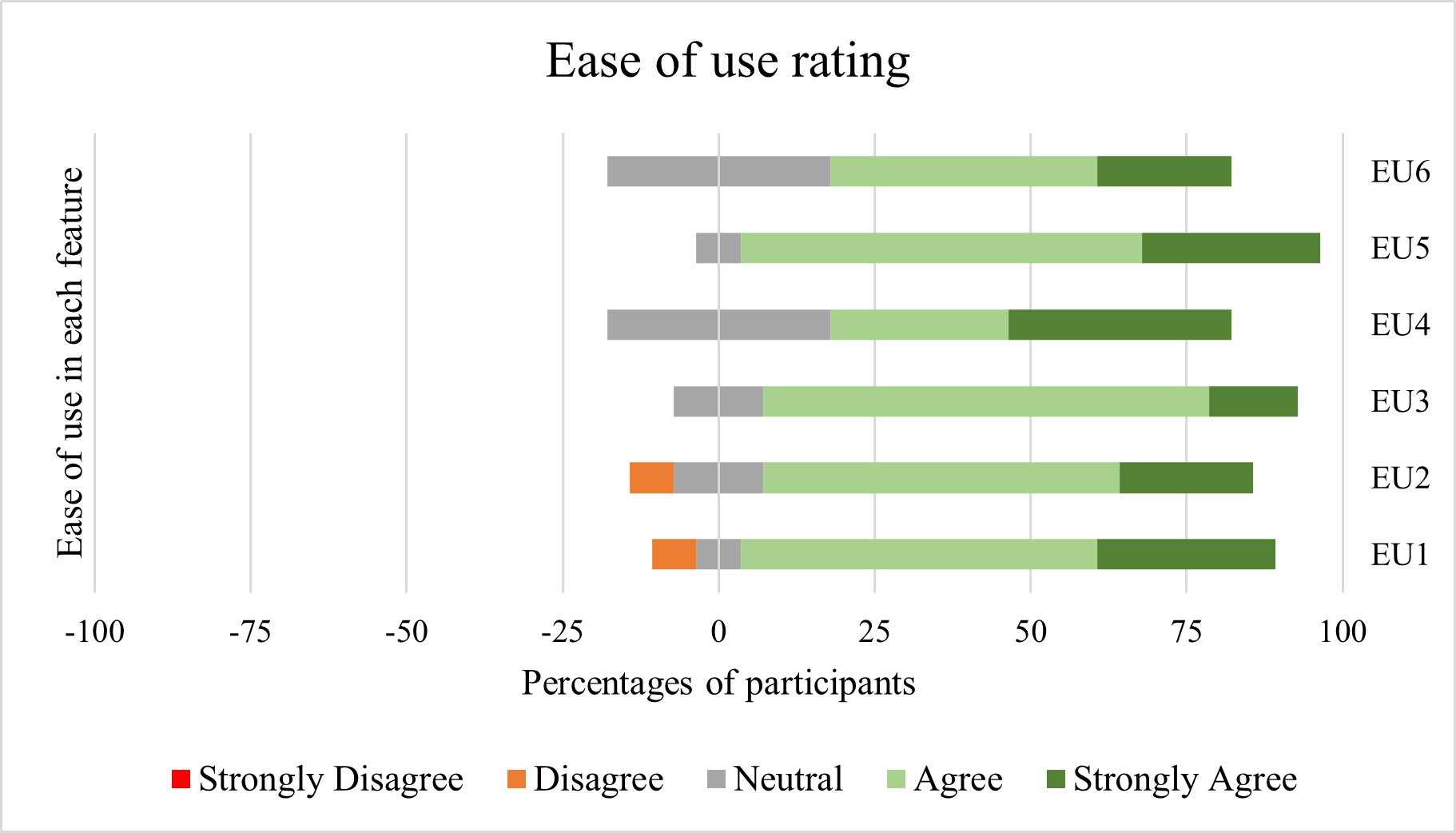}}
        \subfloat[Fun to use]{\includegraphics[width=0.4\textwidth,keepaspectratio]{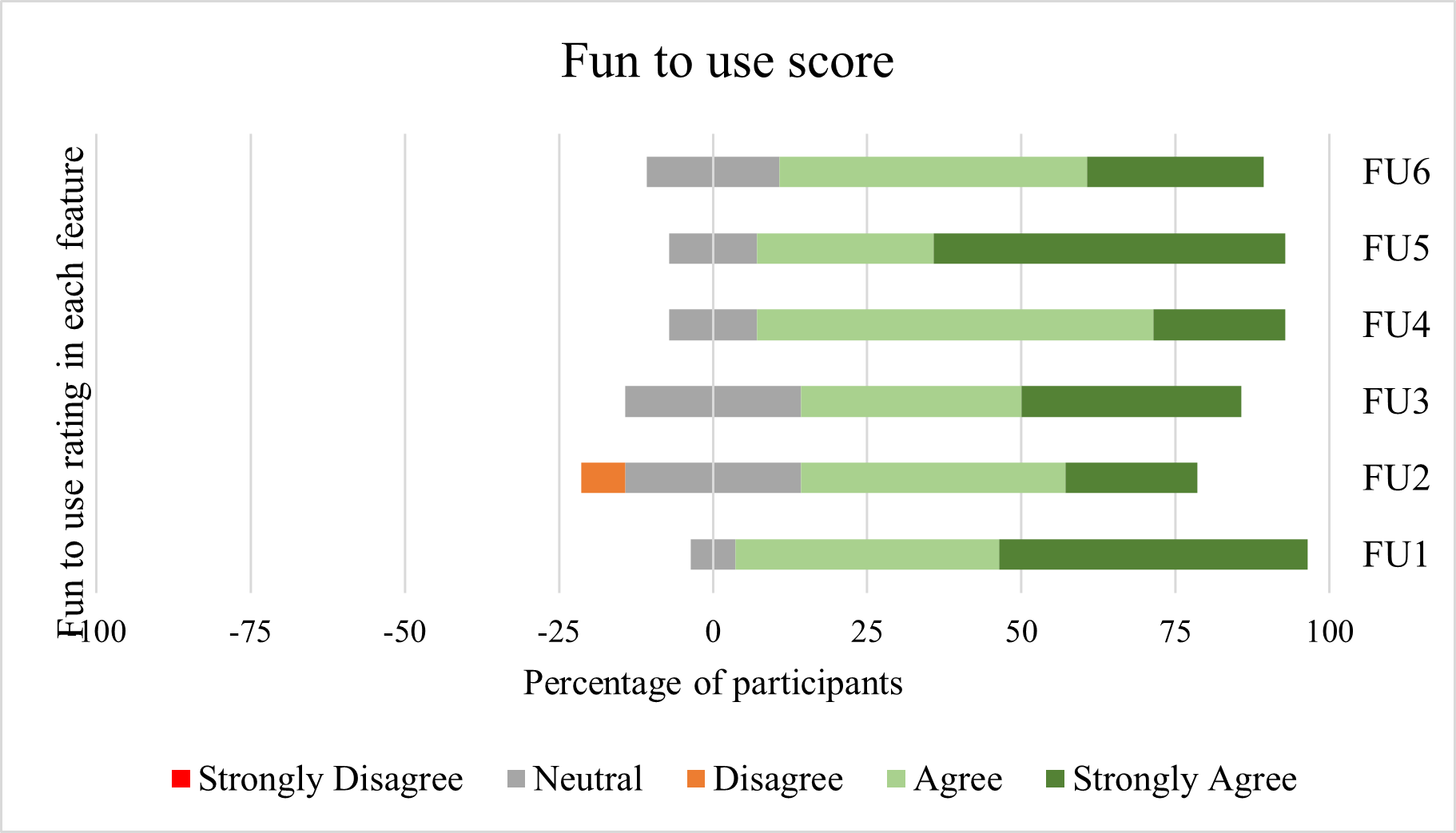}}\\
        \subfloat[Want to use]{\includegraphics[width=0.4\textwidth,keepaspectratio]{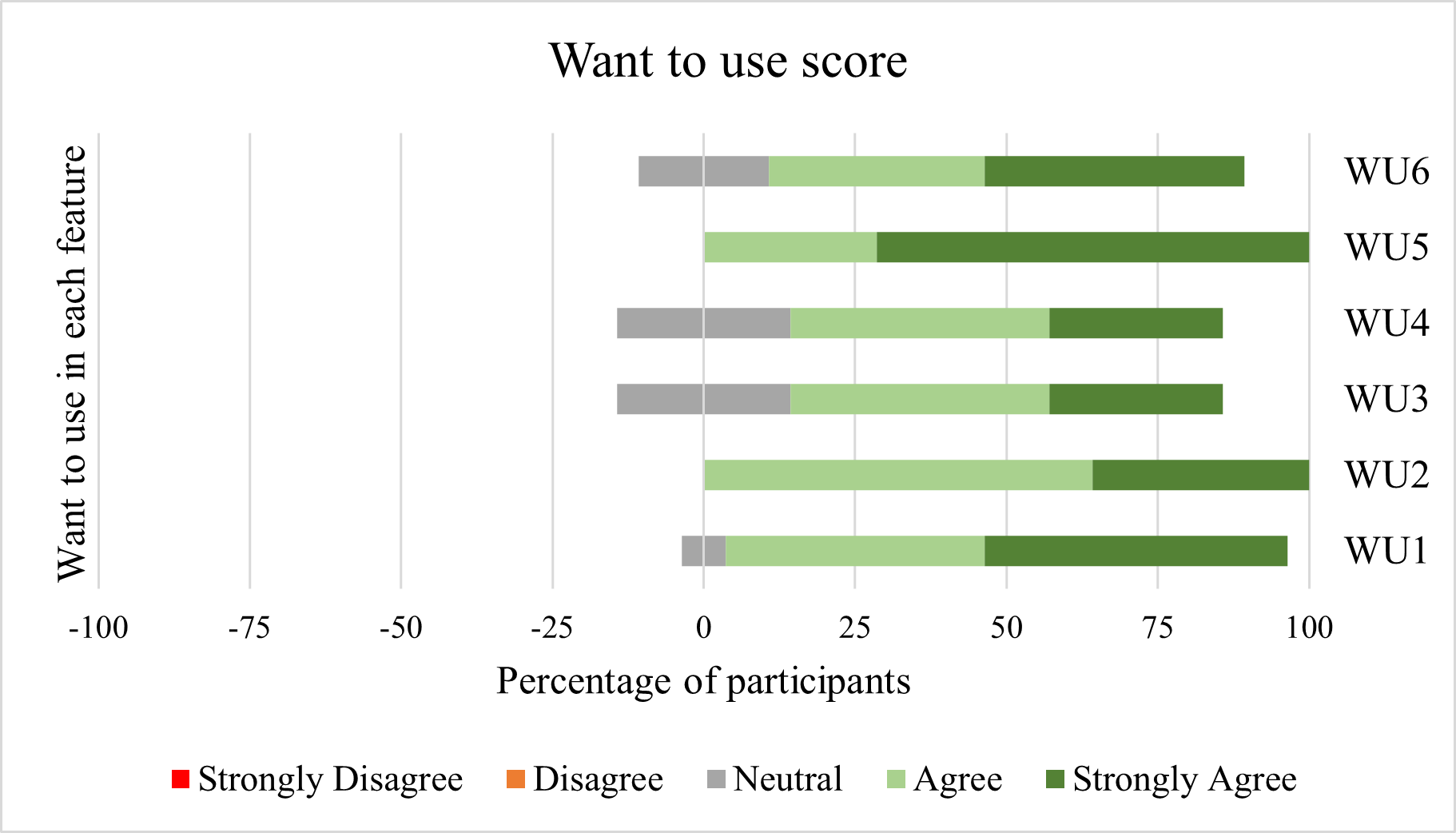}}
        \subfloat[Intuitiveness]{\includegraphics[width=0.4\textwidth,keepaspectratio]{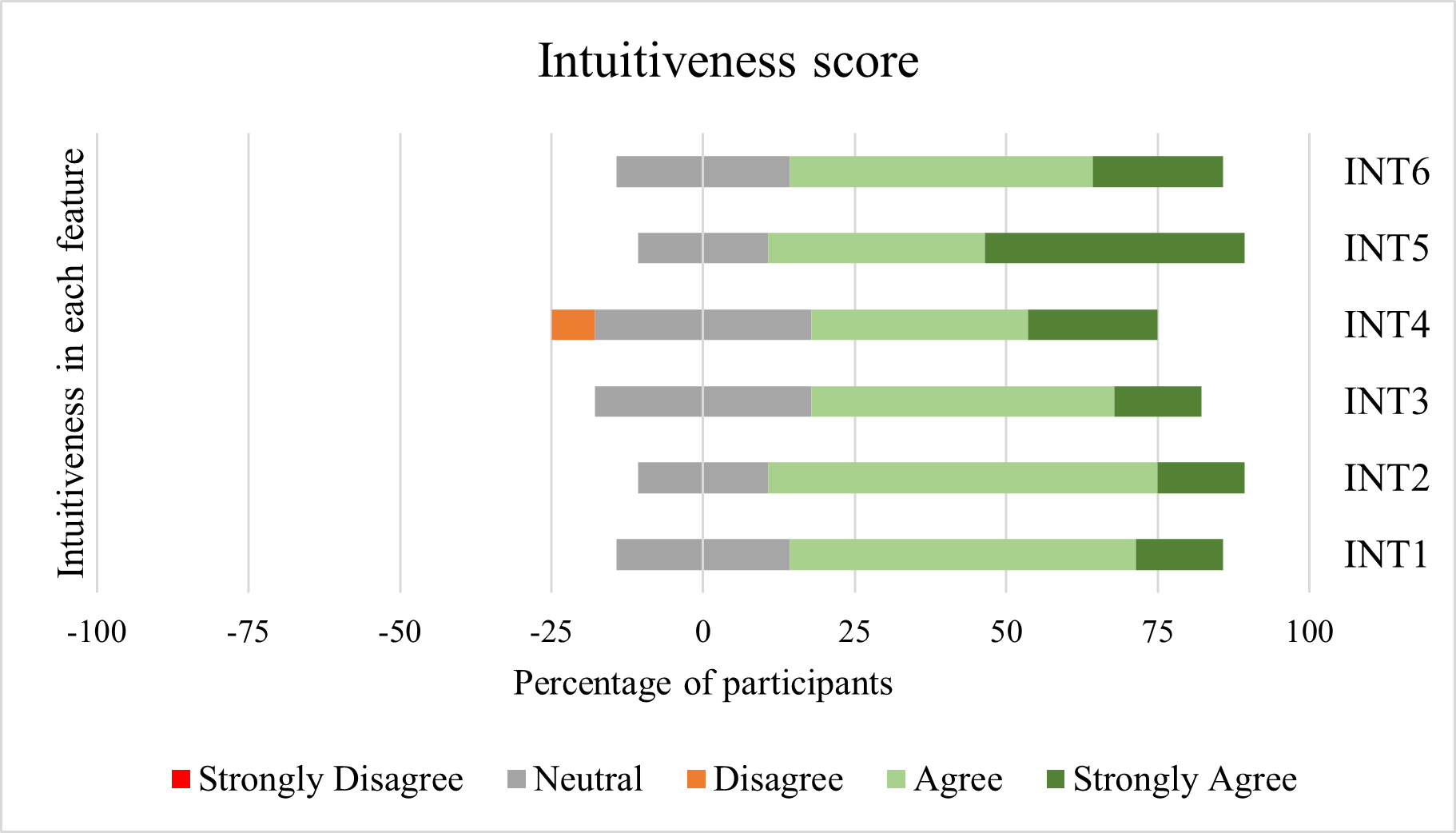}}
    \end{center}
    \caption{Subjective satisfaction of the participants. (Source: authors)}
    \label{Results}
\end{figure}

From Fig. \ref{Results}, it can be seen that the participants have a positive overall satisfactory experience with SIMBA. 7\% of participants expressed dissatisfaction in EU and FU with the sketching feature. For all the remaining features, all the participants showed more than neutral satisfaction. That means the participants did not show negative subjective feelings or displeasure while using the application. The other EU parameters have a minimum score of 3 (Neutral), which shows that the participants had a neutral feeling for EU2, EU4, and EU6. EU3 and EU5 have a minimum score of 4 (Agree). For the parameter that is wanted to be used, the minimum score was 3 (neutral feeling) for all the features. This shows that the participants have a desire for all the features. The fun-to-use parameter's minimum score was 2 (disagree) for FU2. For others, the participants "Agreed" at least that the SIMBA had a fun factor involved in it. The intuitiveness parameter's minimum score was 2 (disagree) for the INT4 feature. For all others, the minimum score was 3. This implies that the participants had at least a "Neutral" feeling about the intuitiveness of the SIMBA. They did not have a negative feeling, which indicates that SIMBA features were intuitive enough to learn and use. All the features created a positive satisfaction, except F2 and F4.

\subsection{Estimation of mental effort}
As mentioned in section 4, the design experiment participants generated design concepts using SIMBA. A think-aloud methodology is used in which the designer verbalises their thought process to externalise the intended motion through the design actions. This design activity is audio and video recorded and analysed using protocol analysis. For analysis, a modified coding scheme based on \cite{Suwa1998} is followed. 

\subsubsection{Data processing}

\begin{table}[]
    \centering
    \caption{Inter-coder reliability results (Source: authors)}
    \label{tab:inter-coder_reliability}
    \begin{tabular}{
        >{\centering\arraybackslash}m{8mm}
        >{\centering\arraybackslash}m{12mm}
        >{\centering\arraybackslash}m{22mm}
        >{\centering\arraybackslash}m{18mm}
        >{\centering\arraybackslash}m{25mm}
    }
    \toprule
    \textbf{S.No.} &
    \textbf{Participants} &
    \textbf{Video duration (seconds)} &
    \textbf{Mental effort (\%)} &
    \textbf{Krippendorff's alpha} \\
    \midrule
    1 & P5 & 1308 & 5.26  & 0.78 \\
    2 & P8 & 488  & 8.45  & 0.68 \\
    3 & P2 & 1006 & 20.0  & 0.97 \\
    4 & P7 & 855  & 20.9  & 0.92 \\
    \bottomrule
    \end{tabular}
\end{table}
Sketching is predominantly used as a tool to visualise concepts. Sketching activity can be considered as a series of discrete design actions. Each action corresponds to a specific design intent. So, it is important to categorise and highlight those actions that are related to motion exploration. Through this categorisation, data is processed to estimate the mental effort involved in motion exploration. 

The raw data in the form of audio-video was transcribed and refined using Adobe Premiere Pro $^{TM}$ to transcribe and refine the verbalisations. The final audio-video is processed in the form shown in Fig. \ref{fig:exp_setup_AND_dataProcess}(b). The processed data from the coders is used to estimate the mental effort of one female and three male participants. The processed videos with subtitles were given to three coders. The coding scheme mentioned in \cite{ramana2020designers} is used in this work. The coders were not given the final outcome (i.e. estimation of mental effort) to keep the results unbiased. The inter-coder reliability of the data is shown in Table \ref{tab:inter-coder_reliability}. The inter-reliability is in the range of [0.68, 0.97], which is above the acceptable standards \cite{blessing2009drm}, i.e. 70\%.

\subsubsection{Quantitative data analysis}
\begin{table}[h]
\centering
\caption{Comparison of mental effort between traditional sketching medium and SIMBA. (Source: authors)}
\label{tab:compare_trad_SIMBA}
\begin{tabular}{>{\centering\arraybackslash}m{45mm} >{\centering\arraybackslash}m{15mm} >{\centering\arraybackslash}m{15mm}}
\toprule
\textbf{Parameter} & \textbf{Traditional} & \textbf{SIMBA} \\
\midrule
Minimum (in \%) & 51.4 & 5.26 \\
Maximum (in \%) & 69.33 & 20.9 \\
Mean (in \%) & 58.89 & 13.64 \\
Standard Deviation ($\sigma$) & 7.66 & 7.96 \\
Standard Error Mean (SEM) & 3.83 & 3.98 \\
Median (in \%) & 57.405 & 14.2 \\
Q1 (in \%) & 54.51 & 7.62 \\
Q2 (in \%) & 57.405 & 14.2 \\
Q3 (in \%) & 61.78 & 20.23 \\
IQR (in \%) & 7.27 & 12.61 \\
Average time spent (seconds) & 333.5 & 914.25 \\
\bottomrule
\end{tabular}
\end{table}

\begin{figure}
    \centering
    \includegraphics[width = 0.6\textwidth, keepaspectratio]{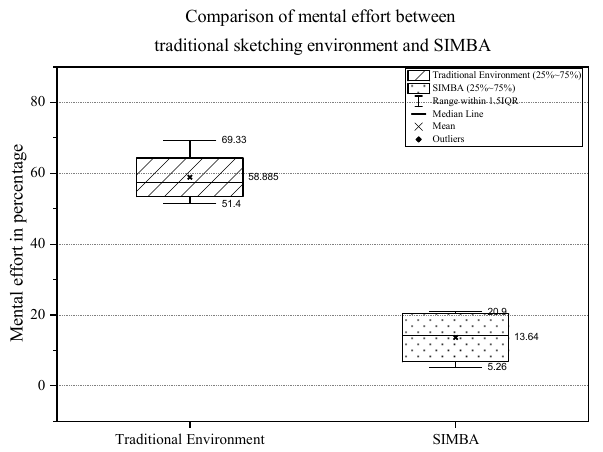}
    \caption{Box plots of mental effort (DV) in motion exploration between traditional sketching medium and SIMBA. (Source: authors)}
    \label{fig:mentaleffort_comparison}
\end{figure}

The results in table \ref{tab:inter-coder_reliability} show that the mental effort in motion exploration ranges from 5.26 to 20.9\%. The average mental effort estimated in this study is 13.625\% (see table \ref{tab:compare_trad_SIMBA}, which shows a decrease of 77\% in mental effort involved in motion exploration compared to that of the traditional sketching environment \cite{ramana2020designers}. This shows that the digital sketching environment's characterisation has effectively decreased the mental effort in motion exploration. This is represented in the form of box plots as shown in Fig. \ref{fig:mentaleffort_comparison}. There are no outliers in the data. The average time spent in traditional is 333.5 seconds, and in SIMBA it is 914.25 seconds. Even though the time spent was more in SIMBA, the mental effort involved in motion exploration was less. This shows that even though SIMBA was more engaging than the traditional sketching medium (evident from the subjective satisfaction), the mental effort was less intense in SIMBA. The sample size of the current study was only four, which cannot be used to prove its significance. The low sample size is due to difficulties in protocol analysis, i.e. recruiting coders, training them, coding the data and checking the inter-coder reliabilities. Each coder usually took 1-1.5 hours on average to code for an 8-10 minute sketching video. So, the process is laborious and tedious. Coding time is proportional to sketching video. For a traditional sketching environment, to make the protocol analysis easier, there could be an automated way of categorising the design actions as per the codes from the recorded audio-video. Hence, the sample size is very low in protocol analysis. Due to the low sample size, the results cannot be shown as statistically significant. For this, bootstrap algorithms \cite{freedman2007statistics} are employed on the current sample to show their statistical significance. 

The effect of SIMBA on the mental effort made by the designer involved in motion exploration is to be tested. In other words, when SIMBA is introduced to designers, is there any change in mental effort during motion exploration when compared to that in traditional sketching media? The mental effort in motion exploration in traditional and SIMBA environments is characterised by the mean of the mental effort of the sample of participants. Let $\mu_{trad}$ and $\mu_{SIMBA}$ be the mean mental effort of the participants estimated during design activity in the traditional and SIMBA environment while motion exploration, respectively. $\mu_{trad}$ - $\mu_{SIMBA}$ is also called mean differnce. To prove the significance of SIMBA, the following hypothesis is formulated

\begin{itemize}
    \item Null Hypothesis: $\mu_{trad} - \mu_{SIMBA}$ = 0
    \item Alternate Hypothesis: $\mu_{trad} - \mu_{SIMBA}$ $>$ 0
\end{itemize}

The null hypothesis states that there is no difference in mental effort in motion exploration when a designer uses traditional and SIMBA sketching environments. The alternate hypothesis states that there is a higher mental effort in a traditional sketching environment compared to the SIMBA sketching environment. The participants for traditional and SIMBA sketching activities were different, so the samples are independent. There are two independent variables (IV), i.e. traditional and SIMBA environment, and there is one dependent variable (DV), i.e. the mental effort involved in motion exploration. The IVs are nominal data, and DV is ratio data. The relation between the IVs and DVs is shown as box plots in Fig. \ref{fig:mentaleffort_comparison}. To test for significance, significance level ($\alpha$) and confidence interval were 0.05 and 95\% respectively. For this, an independent t-test can be used. To evaluate the statistical significance, IBM SPSS is used. 

\begin{table}[ht]
\centering
\caption{Independent Samples Test}
\label{tab:indpendent_samples_test}
\begin{tabular}{
    @{}>{\raggedright\arraybackslash}m{1.8cm}  
    r  
    r  
    >{\centering\arraybackslash}m{1.3cm}  
    >{\centering\arraybackslash}m{1.2cm}  
    >{\centering\arraybackslash}m{1.2cm}  
    >{\centering\arraybackslash}m{1.2cm}  
    >{\centering\arraybackslash}m{1.2cm}  
@{}}
\toprule
\textbf{Mental Effort} & \textbf{t} & \textbf{df} & 
\begin{tabular}[c]{@{}c@{}}\textbf{Sig.}\\\textbf{(1-tailed)}\end{tabular} &
\begin{tabular}[c]{@{}c@{}}\textbf{Sig.}\\\textbf{(2-tailed)}\end{tabular} &
\begin{tabular}[c]{@{}c@{}}\textbf{Mean}\\\textbf{Diff.}\end{tabular} &
\begin{tabular}[c]{@{}c@{}}\textbf{SEM}\\\textbf{}\end{tabular} &
\begin{tabular}[c]{@{}c@{}}\textbf{95\% CI}\end{tabular} \\
\midrule
Equal variances assumed     & 8.182 & 6.000  & $<$.001 & $<$.001 & 45.232 & 5.527 & [31.706, 58.758] \\
Equal variances not assumed & 8.182 & 5.991  & $<$.001 & $<$.001 & 45.232 & 5.527 & [31.701, 58.763] \\
\bottomrule
\end{tabular}

\vspace{1em}
\footnotesize\textsuperscript{a} No statistics are computed for one or more split files.
\end{table}

Table \ref{tab:indpendent_samples_test} shows results for the independent samples t-test. The mean difference between the samples is 45.232 with a SEM = 5.527, CI = [31.706, 58.758]. The test yielded a p-value $<$0, which shows that there is enough evidence to reject the null hypothesis. This implies that there is a difference between the traditional and SIMBA environments. The t-statistic = 8.182, which is $>$ 0, shows that the mental effort in the traditional environment is greater than the SIMBA environment during motion exploration.

\begin{table}[ht]
\centering
\caption{Group Statistics}
\label{tab:groupstatistics}
\begin{tabular}{@{}llr|rrrr@{}}
\toprule
\textbf{Group} & \textbf{Measure} & \textbf{Statistic} & \textbf{Bias\textsuperscript{c}} & \textbf{Std. Error\textsuperscript{c}} & \textbf{Lower\textsuperscript{c}} & \textbf{Upper\textsuperscript{c}} \\
\midrule
\multirow{4}{*}{Traditional} 
& N & 4 & -- & -- & -- & -- \\
& $\mu_{trad}$ & 58.885 & -0.0351\textsuperscript{d} & 3.5125\textsuperscript{d} & 53.475\textsuperscript{d} & 64.736\textsuperscript{d} \\
& $\sigma$ & 7.667 & -1.558\textsuperscript{f} & 2.829\textsuperscript{f} & 3.745\textsuperscript{f} & 8.291\textsuperscript{f} \\
& SEM & 3.833 & -- & -- & -- & -- \\
\midrule
\multirow{4}{*}{SIMBA} 
& N & 4 & -- & -- & -- & -- \\
& $\mu_{SIMBA}$ & 13.653 & -0.0121\textsuperscript{e} & 3.77\textsuperscript{e} & 6.323\textsuperscript{e} & 20.4500\textsuperscript{e} \\
& $\sigma$ & 7.964 & -1.623\textsuperscript{g} & 2.666\textsuperscript{g} & 6.668\textsuperscript{g} & 7.621\textsuperscript{g} \\
& SEM & 3.982 & -- & -- & -- & -- \\
\bottomrule
\end{tabular}

\vspace{1em}
\begin{minipage}{0.95\linewidth}
\footnotesize
\textsuperscript{c} Unless otherwise noted, bootstrap results are based on 2000 bootstrap samples. \\
\textsuperscript{d} Based on 1991 samples. \quad
\textsuperscript{e} Based on 1994 samples. \quad
\textsuperscript{f} Based on 1945 samples. \\
\textsuperscript{g} Based on 1928 samples.
\end{minipage}
\end{table}

\begin{table}[ht]
\centering
\caption{Bootstrap for Independent Samples Test}
\label{tab:bootstrap_results}
\begin{tabular}{
    @{}>{\centering\arraybackslash}m{2cm}  
    >{\centering\arraybackslash}m{1.5cm}  
    >{\centering\arraybackslash}m{1.5cm}  
    >{\centering\arraybackslash}m{1.5cm}  
    >{\centering\arraybackslash}m{1.5cm}  
    >{\centering\arraybackslash}m{2cm}  
    >{\centering\arraybackslash}m{2cm}  
@{}}
\toprule
\textbf{Mental Effort} &
\textbf{\begin{tabular}[c]{@{}c@{}}Mean\\Difference\end{tabular}} &
\textbf{Bias} &
\textbf{\begin{tabular}[c]{@{}c@{}}SEM\end{tabular}} &
\textbf{\begin{tabular}[c]{@{}c@{}}Sig.\\(2-tailed)\end{tabular}} &
\textbf{\begin{tabular}[c]{@{}c@{}}BCa 95\% CI\end{tabular}} \\
\midrule
Equal variances assumed     & 45.232 & -0.0068\textsuperscript{b} & 5.143\textsuperscript{b} & .005\textsuperscript{b} & [35.761, 55.715]\textsuperscript{b} \\
Equal variances not assumed & 45.232 & -0.0068\textsuperscript{b} & 5.143\textsuperscript{b} & .005\textsuperscript{b} & [35.761, 55.715]\textsuperscript{b} \\
\bottomrule
\end{tabular}

\vspace{1em}
\footnotesize\textsuperscript{a} Unless otherwise noted, bootstrap results are based on 2000 bootstrap samples.\\
\textsuperscript{b} Based on 1984 samples.
\end{table}

For bootstrapping, 2000 samples were considered with bias-corrected and accelerated (BCa). Table \ref{tab:groupstatistics} shows group statistics of 2000 samples. Note that the mean, SD and SEM did not change much compared to the original data in table \ref{tab:indpendent_samples_test}. Table \ref{tab:bootstrap_results} shows the results of the statistical test performed on the 2000 samples. The mean difference between the mental effort in traditional and SIMBA environments is 45.232\% (bias = -0.00687, SEM = 5.143, CI=[35.761, 55.715]).  The t-statistic can be estimated from the data, i.e. 45.232/5.143 = 8.79 (t-value = Mean difference/SEM \cite{freedman2007statistics}), which is close to the original t-statistic ($\approx$ 8.182) shown in Table \ref{tab:indpendent_samples_test}. The significance value is estimated to be 0.005 for 2-tailed test. As the p-value is less than the level of significance, there is enough evidence to reject the null hypothesis, which implies that there is a significant difference between mental effort in motion exploration in the traditional and SIMBA environments. Specifically, the mental effort in motion exploration in a traditional environment is higher than in SIMBA environments after bootstrapping. 

\subsection{Qualitative analysis of SIMBA}
\begin{figure}[!h]
    \begin{center}
        \subfloat[P2]{\includegraphics[width=0.25\textwidth,keepaspectratio]{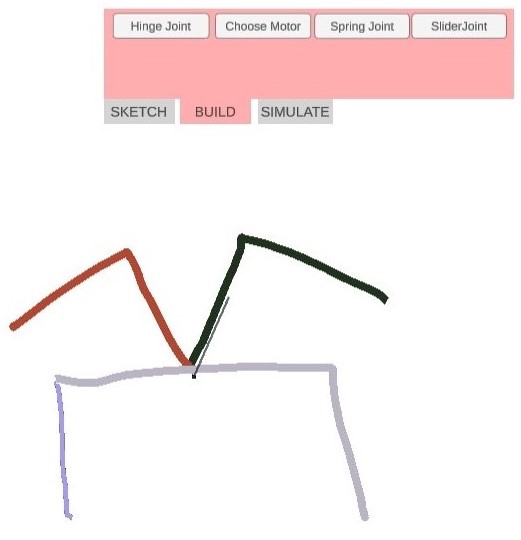}}
        \subfloat[P4]{\includegraphics[width=0.36\textwidth,keepaspectratio]{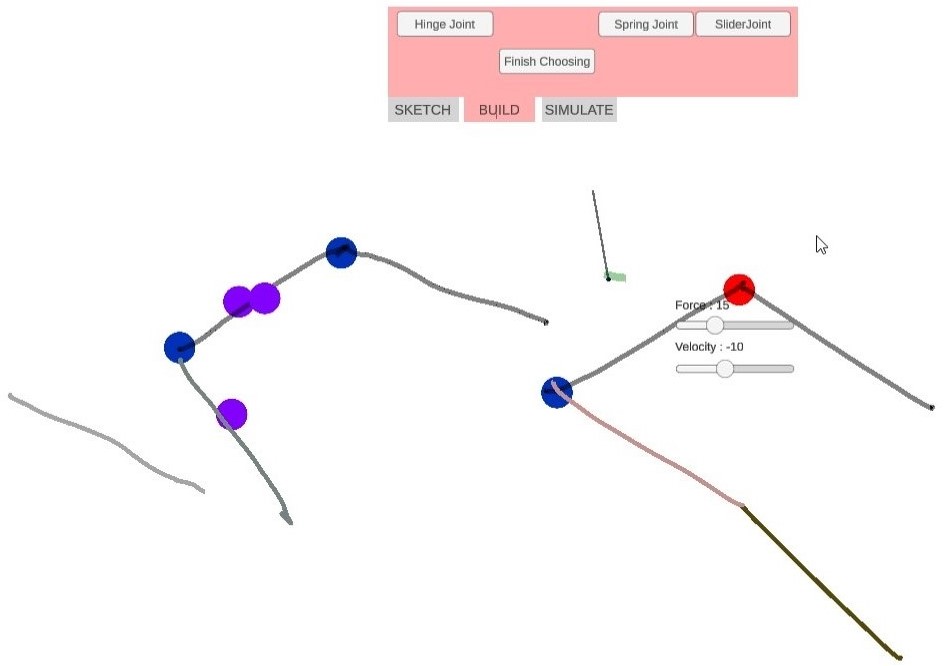}} \hspace{1cm}
        \subfloat[P5]{\includegraphics[width=0.3\textwidth,keepaspectratio]{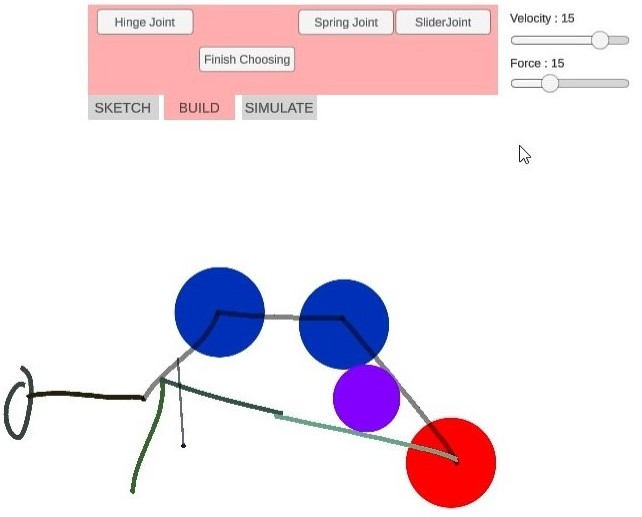}}\\
        \subfloat[P6]{\includegraphics[width=0.28\textwidth,keepaspectratio]{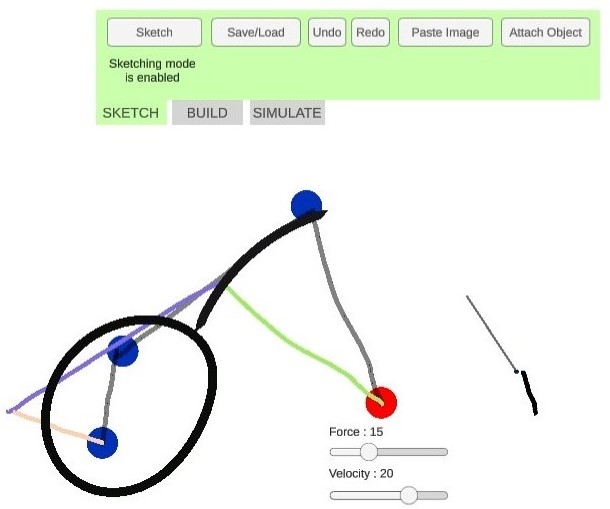}} \hspace{1cm}
        \subfloat[P7]{\includegraphics[width=0.25\textwidth,keepaspectratio]{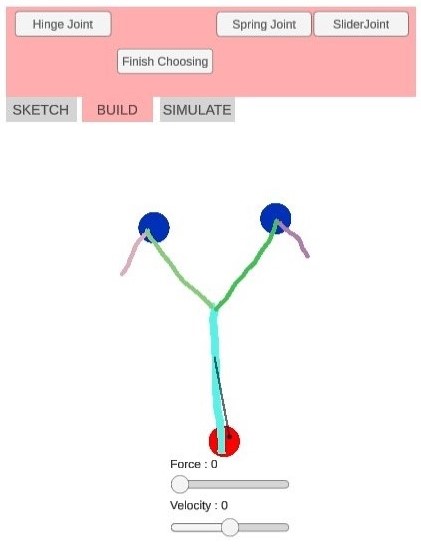}} \hspace{1cm}
        \subfloat[P8]{\includegraphics[width=0.28\textwidth,keepaspectratio]{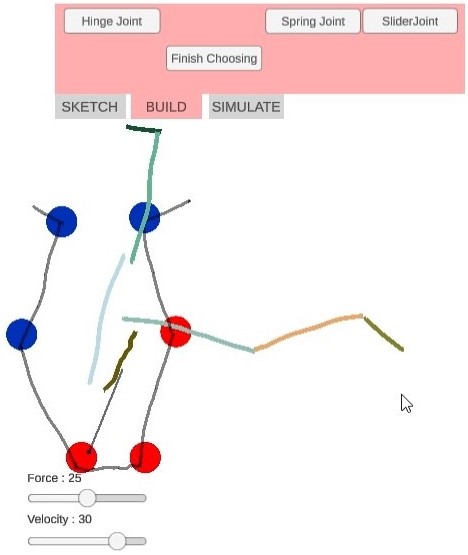}}\\
        \subfloat[P9]{\includegraphics[width=0.28\textwidth,keepaspectratio]{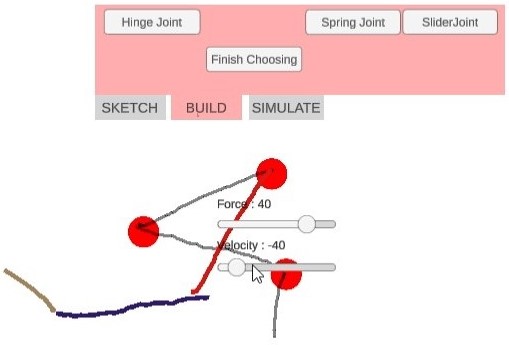}} \hspace{1cm}
        \subfloat[P11]{\includegraphics[width=0.28\textwidth,keepaspectratio]{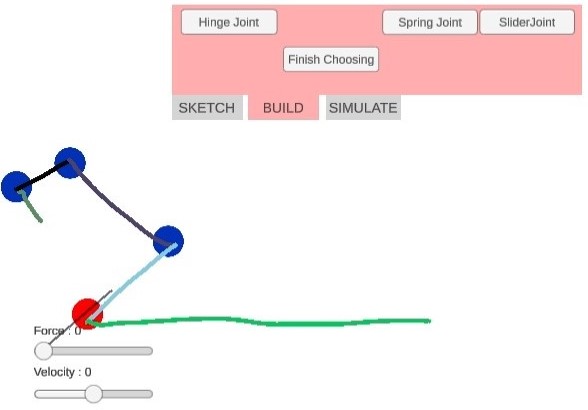}}\hspace{1cm}
        \subfloat[P12]{\includegraphics[width=0.25\textwidth,keepaspectratio]{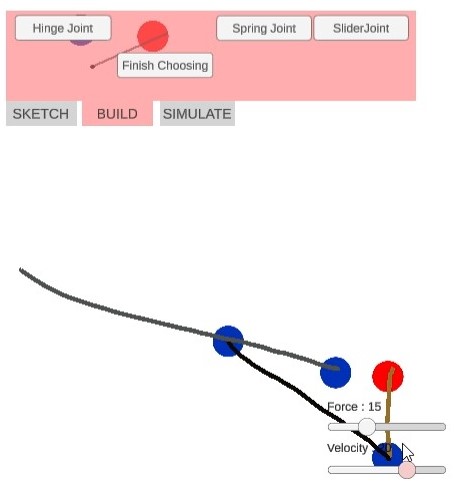}}
    \end{center}
    \caption{Some of the design outcomes from the experiment. (Source: authors)}
    \label{design_outcomes}
\end{figure}
The sketching activity of the participants was studied to determine their challenges that can be divided into the categories as per Fig. \ref{flow_chart} at different stages, viz., synthesis, analysis, and validation. During synthesis, the participants could draw the elements and establish a relation between them in the articulated product concept seamlessly. Most of the participants used a matchstick figure as their inspiration to generate a concept (see Fig. \ref{design_outcomes}(a) and (e)). In reality, the neck and shoulder are not joined at a point. Technically, the abstraction itself was not up to the mark. This led to establishing wrong or inappropriate motion constraints for the sub-components. Participant P2 reported difficulty in getting used to the stylus of the haptic device. Participant P5 reported that the smoothening of the strokes could be a nice aesthetic visual feature while sketching in SIMBA. But as per literature \cite{Schon1992}, once the naturality of the strokes is manipulated, it will affect the cognition of the designer. Hence, this feature could be optional in the future.

    \begin{figure}[!h]
    \begin{center}
        \subfloat[Embodiment of P6 concept]{\includegraphics[width=0.9\textwidth,keepaspectratio]{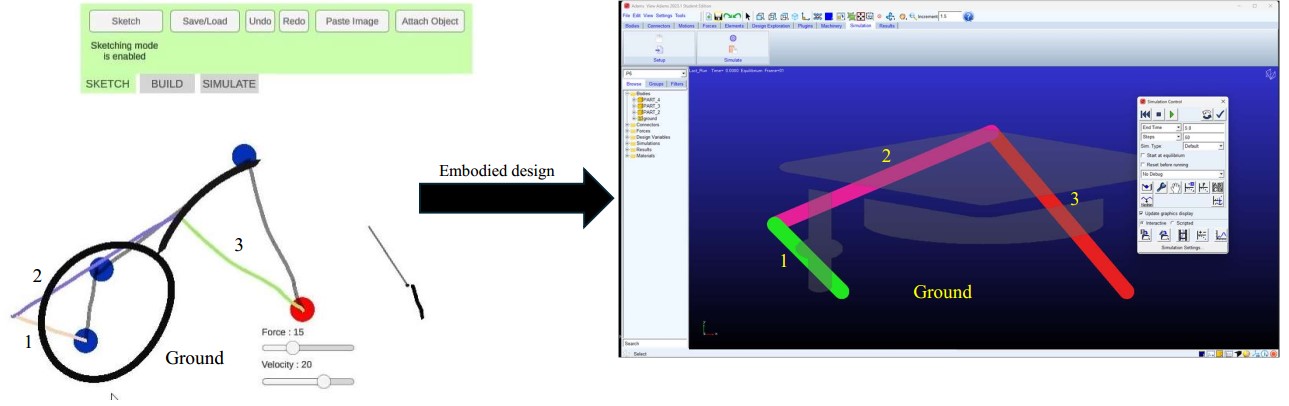}}\\
        \subfloat[Embodiment of P7 concept]{\includegraphics[width=0.9\textwidth,keepaspectratio]{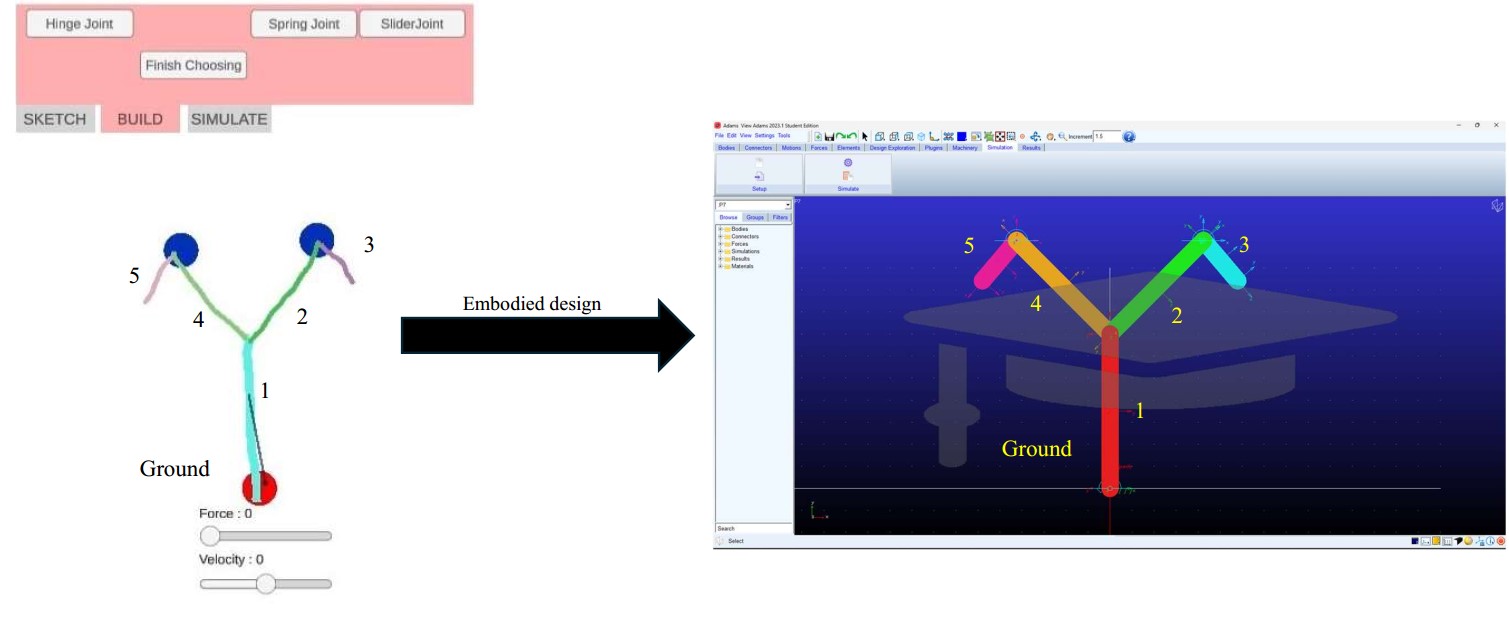}}
    \end{center}
    \caption{Embodiment of the SIMBA models in ADAMS. (Source: authors)}
    \label{Embodiment}
    \end{figure}

During the design analysis phase, participants faced a technical challenge. They failed to notice the fundamental function of the motor (a device that converts electrical energy to mechanical energy in the form of simple motions), and they were expecting the software application (SIMBA) to give their own required characteristic motion. Some participants asked for rocking (or to-and-fro motion) of a particular sub-component. Methods mentioned in \cite{hall1986kinematics} can be used to synthesise the rocking motion of a 4-bar mechanism. Perhaps the lack of this technical knowledge is somewhat blamed on the software that has to provide such an option. 

During the validation phase, the articulated product concepts generated by the participants did not meet the criteria mentioned in the design requirements, except for participant P6. This articulated product differed from the rest because it used a 4-bar mechanism to get a rocking motion. The solution made by P4 seems most reasonable. To evaluate, the current concepts have been modelled in ADAMS software as shown in Fig.\ref{Embodiment}. The abstract skeleton structure is converted into a digital model in ADAMS. As such, the design outcomes do not work if implemented "as they are", and an extra effort is to be made to make them work. 

\section{Discussion}
\label{discussion}
The present work focuses on integrating the structural and functional paradigms of conceptual design. It provides unique features for designing structural elements \cite{Tversky2002, Bae2008} such as form, colour, texture, etc, integrated with the functional paradigm of motion synthesis \cite{Alvarado2001, PurwarMotion_Gen, FEAsyJournal2017}. The ideas mentioned in \cite{ball2019advancing} about the concept of structurally and functionally oriented sketches are combined in the current work. The "naturalness" of the traditional sketching activity is characterised and integrated in SIMBA to simulate the articulated product concept sketches, thus reducing the mental effort of the designer.

This work opens new possibilities for supporting designers' freeform sketching in the early phases. The literature on sketch analysis \cite{Onkar2010} identify the design intent of the freeform sketches that can be enhanced with similar interfaces like SIMBA for motion behaviour analysis. Designers are also interested in other behavioural analyses, such as structural, thermal, and fluid interactions, but the work presented here is limited only to the motion behaviour simulation. 

As noted by \cite{wetzel2009automated}, the absence of embedded motion-related knowledge made it challenging for designers to generate and develop concepts effectively. The concept generated by participants can be validated by making physical prototypes or by modelling in a simulation software.  ADAMS$^{TM}$ is one of the commercially available software used to simulate the motion of product concepts. 
 
 Fig. \ref{ex_1_simulation} (bass drum mechanism sketch) shows snapshots captured at discrete time intervals to illustrate the simulation of the explored two product concepts. Moving components are highlighted in blue, while static components are shown in black in Fig. \ref{ex_1_simulation}(a). The designer can explore how much the crank has to rotate to move the drumstick as close as it is to the drum. Fig \ref{ex_1_simulation}(d)-(i) shows the relationship between the drumstick and foot pedal rotations (in both clockwise and counterclockwise directions) across various phases. Such inferences are challenging to obtain from static sketches. Thus, it is believed to be affecting the mental effort of the designers. Further investigations could lead to the enhancement of cognitive capabilities built into the tool itself. 

Some of the features, like adjusting the joint location, designers can also refine the motion to meet the requirements. In the SIMBA design experiment, the feature was used to explore variations of concepts. Even though the sketching application has achieved its goal of motion visualisation of a product concept without deviating from the sketching activity, its response concerning each factor of subjective satisfaction was positive. This shows that there is a level of satisfaction among the participants in using the sketching application to visualise the motion of the product concept. This shows that even though the primitive geometries have not been used to represent the links or rigid bodies as mentioned in \cite{eicholtz_kara}, \cite{chase2013}, \cite{Cheng2005}, \cite{KaustubhMcCarthy2015},  \cite{PurwarMotion_Gen}, \cite{mauskar}, \cite{autoSketch_Kecksmity}, \cite{furlong1999} and \cite{Kihonge2002} the designers are able to visualize the motion without any discomfort or difficulty. This may be because the concept ideas came from their minds, and it was easy for them to relate to the product concept in the sketching application. The difficulty or discomfort may increase if they are asked to read or re-interpret the designs made by others. In this aspect, the criteria for the re-interpretation of one designer should match the intention of the idea of the original owner of the sketch. Thorough usability (effectiveness and efficiency) testing in the future may reveal further feature improvements. Some features, such as path tracing, joint manipulation and multiple mechanisms, are reported as "beneficial" features for sketching articulated product concepts by participants. 

\section{Conclusion and Future work}
An interactive software tool, SIMBA, enables the synthesis, analysis and evaluation of articulated product concepts. The application incorporated the aspects of the sketching activity into a digital sketching medium so that the users do not digress away from the sketching activity and simulate their concepts for motion validation. Two types of evaluations were performed, i.e., subjective satisfaction of the interface by the designers and the estimation of the designers' mental effort. 
  
Subjective satisfaction with the SIMBA was measured with respect to four parameters: ease-of-use, want-to-use, fun-to-use and intuitiveness. The overall results say that the software was ‘liked’ by the designers for its features that involved sketching, building mechanisms, and simulating them. From this work, it can be concluded that the sketch-based interface for the ideation of articulated product concepts greatly reduces the mental effort of the designers (77\% reduction in mental effort), while motion exploration, compared to a traditional sketching environment. Thus, SIMBA greatly enhances designers' learning of motion exploration in articulated product concept sketches. 

This work opens up a new paradigm of integrated concept development wherein designers can switch between fuzzy concept sketching phases at the same time, having a crisply defined simulation in a concurrent interface. This is supposed to facilitate faster product development. In addition to enhancing the SIMBA interface with features for other kinematic joints, the power of simulation can be extended to the 3D environment \cite{onkar2016controlled}. Combining with virtual/augmented reality environment, the interface can facilitate realistic and immersive feedback. Moreover, if SIMBA is enhanced with haptic feedback, it can accelerate the product development by providing a realistic experience.

\section*{Supplementary information}
The following supplementary materials are provided along with this article. These are also available in public repository 
\begin{verbatim}
https://github.com/gkramana8687/IJIDeM_SupplementaryMaterial.git
\end{verbatim}

\begin{itemize}
    \item SIMBA software. (Folder name: SIMBA.zip)
    \item Video recording of the sketching activities of four participants. (Folder name: Processed videos for transcription)
    \item Spreadsheets that contain the protocol analysis of the four participants. (Coding\_IJDeM.zip)
    \item Screen recording of the simulations mentioned in the paper. These are for Fig. \ref{ex_1_simulation}, Fig. \ref{TracingMechanism} and Fig. \ref{multiple_mechanisms}. (Videos related to the article)
\end{itemize}

\section*{Acknowledgements}
Supported by the Department of Science and Technology (DST),  Science and Engineering Research Board (SERB) (File no. CRG/2020/005334).

\section*{Declarations}
Generative AI is not used to write any part of the manuscripts.

\bibliography{Arxiv_IJDEM}
\end{document}